\documentclass[a4paper,onecolumn,11pt,accepted=2026-10-06]{quantumarticle}
\pdfoutput=1
\usepackage[utf8]{inputenc}
\usepackage[english]{babel}
\usepackage[T1]{fontenc}
\usepackage{amsmath}
\usepackage{hyperref}
\usepackage{tikz}
\usepackage{lipsum}

\usepackage{graphicx, dcolumn, epsfig, float, algorithm, algorithmic, lipsum, mathrsfs, eufrak, tikz, soul, trsym, xcolor,  orcidlink}
\usetikzlibrary{quantikz}
\hypersetup{colorlinks=true,citecolor=blue,linkcolor=blue, urlcolor=blue}
\usepackage[bbgreekl]{mathbbol}
\usepackage{mathrsfs}
\usepackage{hyperref}
\usepackage{comment}
\usepackage{tabularx}
\usepackage{array}
\usepackage{adjustbox}
\usepackage{amssymb}

\usepackage{mathrsfs}
\usepackage{eufrak}
  
\newcommand{\vect}[1]{\boldsymbol{\mathbf{#1}}}

\def\bra#1{\mathinner{\langle{#1}|}}
\def\ket#1{\mathinner{|{#1}\rangle}}
\def\inner#1#2{\mathinner{\langle{#1}|{#2}\rangle}}
\def\braket#1{\mathinner{\langle{#1}\rangle}}
\def\sandwich#1#2#3{\mathinner{\langle{#1}|{#2}|{#3}\rangle}}

\def\bbra#1{\mathinner{\langle\hspace{-0.75mm}\langle{#1}|}}
\def\kket#1{\mathinner{|{#1}\rangle\hspace{-0.75mm}\rangle}}
\def\iinner#1#2{\mathinner{\langle\hspace{-0.75mm}\langle{#1}|{#2}\rangle\hspace{-0.75mm}\rangle}}

\def\Tr{\mathrm{Tr}}

\def\Om{\Omega}

\def\ep{\epsilon}
\def\al{\alpha}

\usepackage[T3,T1]{fontenc}
\DeclareSymbolFont{tipa}{T3}{cmr}{m}{n}
\DeclareMathAccent{\invbreve}{\mathalpha}{tipa}{16}

\usepackage{accents}
\newlength{\hhatheight}
\newcommand{\hhat}[1]{%
    \settoheight{\hhatheight}{\ensuremath{\hat{#1}}}%
    \addtolength{\hhatheight}{-0.35ex}%
    \hat{\vphantom{\rule{1pt}{\hhatheight}}%
    \smash{\hat{#1}}}
}

\begin{document}
\title{Floquet-ADAPT-VQE: A Quantum Algorithm to Simulate Non-Equilibrium Physics in Periodically Driven Systems}

\author{Abhishek Kumar}
\orcid{0000-0003-3358-7787}
\affiliation{Department of Physics, Virginia Tech, Blacksburg, VA 24061, USA}
\affiliation{Virginia Tech Center for Quantum Information Science and Engineering, Blacksburg, VA 24061, USA}

\author{Karunya Shirali}
\orcid{0000-0002-2006-2343}
\affiliation{Department of Physics, Virginia Tech, Blacksburg, VA 24061, USA}
\affiliation{Virginia Tech Center for Quantum Information Science and Engineering, Blacksburg, VA 24061, USA}

\author{Nicholas J. Mayhall}
\orcid{0000-0002-1312-9781}
\affiliation{Department of Chemistry, Virginia Tech, Blacksburg, VA 24061, USA}
\affiliation{Virginia Tech Center for Quantum Information Science and Engineering, Blacksburg, VA 24061, USA}

\author{Sophia E. Economou}
\orcid{0000-0002-1939-5589}
\affiliation{Department of Physics, Virginia Tech, Blacksburg, VA 24061, USA}
\affiliation{Virginia Tech Center for Quantum Information Science and Engineering, Blacksburg, VA 24061, USA}

\author{Edwin Barnes}
\orcid{0000-0002-0982-9339}
\affiliation{Department of Physics, Virginia Tech, Blacksburg, VA 24061, USA}
\affiliation{Virginia Tech Center for Quantum Information Science and Engineering, Blacksburg, VA 24061, USA}

\maketitle
\begin{abstract}
  Periodically driven quantum systems exhibit many fascinating phenomena absent in equilibrium systems, but their simulation is more challenging than that of static systems. Consequently, quantum simulation of these systems offers greater opportunity for achieving quantum advantage. To build the foundation for simulating time-periodic Hamiltonians, we utilize the Floquet-Hilbert space formalism to transform the Hamiltonian into a time-independent form and provide its Pauli decomposition. Within this formalism, we develop a hybrid quantum-classical algorithm, Floquet-ADAPT-VQE, that produces three key quantities that govern periodically driven quantum dynamics: the quasienergy, the associated Floquet eigenstates, and expectation values of observables with respect to these eigenstates. The latter is achieved by applying a fixed-depth quantum circuit to an initial Floquet eigenstate---an efficient construction made possible by the Floquet-Hilbert space formalism. We also identify a suitable  auxiliary initial state whose squared Floquet energy is independent of the number of auxiliary qubits as well as the driving frequency, which leads to better convergence with fewer ADAPT iterations.  We demonstrate our algorithm by performing numerical simulations on a periodically driven XYZ model with a magnetic field. We also explore potential applications of our algorithm for studying various non-equilibrium phenomena in periodically driven systems. 
\end{abstract}

\section{Introduction}

Non-equilibrium many-body quantum systems, particularly periodically driven systems, exhibit a wide array of fascinating phenomena that do not occur in equilibrium systems. These phenomena include, but are not limited to, non-equilibrium phases such as time crystals \cite{Wilczek_2012, Else2016_TC,Yao_2017_TC,Khemani_2016_TC,  Li_2020_HeisenTC,Barnes_2019_HeisenTC,Kumar_Scar_TC_PRB2025,Rafail_PRB_2023} and Floquet topology \cite{Lindner_2011,Potter_2016_PRX,Kitagawa2010,asboth2014,Vega2019}, both of which lack counterparts in equilibrium systems. In general, quasienergies and Floquet states, which are eigensolutions of Schr\"odinger's equation (up to a phase) for periodically driven Hamiltonians, play a crucial role in the manifestation of these remarkable phenomena. For instance, in time crystals that exhibit period doubling, product states can be expressed as linear combinations of two Floquet states separated by half of the driving frequency in the quasienergy spectrum. Therefore, the system manifests a spontaneous breaking of discrete time translation symmetry when it is initialized in a product state. Additionally, Floquet states can exhibit non-equilibrium topology due to the presence of additional gaps between Floquet zone bands. Broadly speaking, periodic drives offer a means to control the electronic properties of quantum matter \cite{Oka2019FloquetEngineering} and could potentially be utilized to create sensors to detect exotic features of a system \cite{Kumar_PRL_2021}. As a result of these rich phenomena, and many others yet to be fully explored, the study of periodically driven systems has garnered significant attention in recent years, emerging as a major focus for researchers in condensed matter physics, atomic, molecular, and optical physics, as well as quantum computation.

Simulating many-body quantum systems, even in equilibrium, poses significant challenges; the task becomes even more daunting in the case of non-equilibrium, driven, physical systems. The complexity of simulating non-equilibrium physics in periodically driven many-body quantum systems stems from two primary challenges: the exponential growth of the Hilbert space with the system size and the errors introduced by  approximating the time evolution operator (e.g., Trotterization). 
On the other hand, using devices based on quantum mechanics to simulate many-body quantum systems, as suggested by Feynman \cite{Feynman_1982}, has emerged as a promising avenue to study non-equilibrium behavior. Quantum simulators or analog quantum computers have already demonstrated the ability to simulate many-body physics that are intractable on classical computers \cite{Zhang_2017_nature, Bernien_2017_nature,Semeghini_Lukin_Science_2021}. However, these simulators are not universal; they function by mimicking many-body interactions and can only simulate a limited range of many-body systems of interest. 

In contrast, digital quantum computers are universal and have the potential to efficiently simulate any many-body system. While a fully fault-tolerant quantum computer would be an ideal platform to perform these simulations \cite{Abrams_Lyod_1999_PRL}, error correction schemes have not yet been able to reach the stage at which such fault-tolerant quantum computers can be realized. However, currently available Noisy Intermediate-Scale Quantum (NISQ) devices could still be a valuable resource in addressing this challenge \cite{Preskill_2018_NISQ}. In this context, hybrid quantum-classical algorithms like the Variational Quantum Eigensolver (VQE) \cite{Peruzzo_2014_VQE,McClean_2016_NJP} are especially promising. In particular, the Adaptive Derivative-Assembled Problem Tailored Variational Quantum Eigensolver (ADAPT-VQE) \cite{Grimsley_NatCom_2019,Tang_QubitADAPT_2021,Anastasiou_Tetris2024}, holds great promise, as it can generate shallow quantum circuits suited for noisy qubits with limited coherence times. It has also been found to be robust against some optimization problems that plague VQEs \cite{Grimsley_2023_ADAPT_Barren}.

The key challenge is to devise quantum algorithms that can produce the quantities that govern the non-equilibrium dynamics generated by time-periodic Hamiltonians, namely the quasienergy spectrum, the associated Floquet eigenstates, and time-evolving expectation values of observables with respect to Floquet eigenstates. Recently, Fauseweh et al. proposed two hybrid algorithms~\cite{Fauseweh2023}, in the time and frequency domains, to calculate the quasienergies and Floquet eigenstates in periodically driven systems using the variational Hamiltonian ansatz~\cite{Wecker_2015}. Although this work successfully demonstrated the calculation of quasienergies and Floquet eigenstates, both methods have limitations. The time-domain algorithm requires Trotterization for one time period and iterative quantum phase estimation (IQPE)~\cite{Griffiths_IQPEA_1996}, in addition to the usual VQE implementation. This can lead to much deeper circuits and is only effective for short time periods (high frequencies). The frequency-domain algorithm also has some limitations: it uses an auxiliary space with a limited dimension (at most five), and it does not provide any general prescription to obtain the Pauli decomposition of the auxiliary matrices for arbitrary qubit truncation. Consequently, the algorithm can simulate the periodic system accurately only at high driving frequencies. Additionally, a potentially major challenge is ansatz initialization, which can be critical for convergence, since simulations using the variational Hamiltonian ansatz typically require running the optimization multiple times with different randomly chosen initial parameter values and post-selection. Finally, the method does not provide a framework to study the expectation values of operators with respect to Floquet states, which play a crucial role in studying various non-equilibrium phenomena. 

These limitations point to a more fundamental gap in the literature. Broadly speaking, there is no well-established, general-purpose Pauli-string decomposition of the extended Floquet Hamiltonian, which prevents a systematic implementation of both near-term and fault-tolerant quantum algorithms within the Floquet-Hilbert space formalism. Moreover, current approaches lack a comprehensive framework for simulating periodically driven quantum systems at arbitrary driving frequencies without relying on Trotterization or quantum phase estimation (QPE) to compute quasienergies and Floquet states. We address both gaps in this work.

In this paper, we develop a general framework to simulate periodically driven systems in a manner analogous to time-independent Hamiltonians. We employ the Floquet-Hilbert space formalism to map a time-periodic Hamiltonian to an effective time-independent Hamiltonian and provide a systematic method to decompose the auxiliary matrices, and hence the extended Floquet Hamiltonian, into a Pauli basis natively implementable on qubit-based quantum processors. 

Building on this framework, we introduce a hybrid quantum-classical algorithm, Floquet-ADAPT-VQE, to compute quasienergies and prepare extended Floquet states. The algorithm is based on a cost function defined in terms of the square of the shifted, extended Floquet Hamiltonian. Choosing a suitable initial state for the auxiliary space is important to the efficacy of the algorithm due to the dependence of the squared Floquet energy on the number of auxiliary qubits and the driving frequency. We propose choosing this state to be an eigenstate of a diagonal auxiliary matrix with vanishing eigenvalue and prove that, in this case, the squared Floquet energy is independent of both the number of auxiliary qubits and the driving frequency. Additionally, we present a method to compute the full quasienergy spectrum within a Floquet zone.

A particularly important feature of our framework is that it enables access to micromotion dynamics of a Floquet state using fixed-depth quantum circuits. By preparing the extended Floquet state with auxiliary qubits, both time-dependent expectation values of observables and time-periodic Floquet states can be obtained without requiring deep circuits. This contrasts with approaches that work in the original Hilbert space, where circuit depth typically grows with the evolution time.

To illustrate the effectiveness of our algorithm, we classically simulate it for a periodically driven XYZ model with an applied magnetic field. We show that the reference value of the cost function (the squared Floquet energy) of our proposed initial state, unlike that of a random initial state, does not increase even when the number of auxiliary qubits and driving frequency are increased. We demonstrate the effectiveness of our algorithm by testing it across a range of drive parameters. Additionally, we compute the full quasienergy spectrum within the central Floquet zone by tuning a shift parameter in the cost function. Finally, we calculate the expectation values of the net magnetization ($\sum_j Z_{j}$) and the two-point correlation function $\sum_j Z_{j}Z_{j+1}$ using our algorithm.

The paper is organized as follows.
In Sec.~\ref{sec:Primer}, we discuss Floquet theory, its representation in the Floquet-Hilbert space, and its usefulness for NISQ devices. We also provide a brief discussion of ADAPT-VQE. In Sec.~\ref{sec:Floq_ADAPT_Algo}, we show how to construct auxiliary matrices and extended Floquet Hamiltonians in terms of Pauli strings. We also define the cost function based on the squared Floquet Hamiltonian and discuss choices of initial states and operator pools. In addition, we propose a framework to calculate the expectation values of operators with respect to Floquet states on a quantum computer.
In Sec.~\ref{sec:Application}, we present classical simulations of our algorithms applied to a periodically driven XYZ model with an external magnetic field. In Sec.~\ref{sec:Discussion}, we discuss several applications of our algorithm for light-matter interactions in many-body interacting systems, as well as the potential for detecting time crystals using our framework.
Some additional details are given in the appendices.

In this work, we denote an operator in the physical space as $\hat{O}$, in the auxiliary space as $O$, and in the Floquet-Hilbert space as $\hhat{O}$. Similarly, a state in the physical and auxiliary spaces is represented as $\ket{\psi}$, and in Floquet-Hilbert space as $\kket{\psi}$. Throughout this work, we denote an operator and its matrix representation using the same notation. Also, $\mathrm{T}$ in $(\dots,\dots)^\mathrm{T}$ represents the transpose of a matrix throughout this paper. Additionally, we use three kinds of bases for the $2^{n_a}$ dimensional  auxiliary Hilbert space: the Fourier basis $\{\ket{n}\}_{n=-2^{n_a-1}+1}^{2^{n_a-1}}$, the standard vector basis in $\mathbb{R}^{2^{n_a}}$, and the computational basis $\ket{\gamma_1...\gamma_{n_a}}$ where $\gamma_j\in\{\uparrow,\downarrow\}$. For a driven $L$-qubit system, the Floquet-Hilbert space is $\mathbb{R}^{2^{n_a}}\otimes\mathbb{C}^{2^L}$.

\section{Primer on Floquet systems and ADAPT-VQE}\label{sec:Primer}
\subsection{Floquet System }

Floquet systems are governed by a time-periodic Hamiltonian $\hat{H}(t) = \hat{H}(t+T)$, where $T=2\pi/\Omega$ and $\Omega$ is the fundamental frequency of the drive. According to the Floquet theorem \cite{Floquet_1883}, the solutions of the time-dependent Schr\"odinger equation, $i\partial_t\ket{\psi(t)} = \hat{H}(t)\ket{\psi(t)}$ (assuming $\hbar=1),$ take the form $\ket{\psi_{\alpha}(t)} = e^{-i\epsilon_\alpha t} \ket{\phi_{\alpha}(t)}$. The time-periodic component of the solution, $\ket{\phi_\alpha(t)} = \ket{\phi_\alpha(t+T)}$, is known as a Floquet state and satisfies the Floquet-Schr\"odinger equation: 
\begin{equation} \label{eq:FloqSch}
    [\hat H(t) - i\partial_t]\ket{\phi_\alpha(t)} = \epsilon_\alpha \ket{\phi_\alpha(t)}.
\end{equation}
Here, the subscript $\alpha \in \{1, 2, \dots, D\}$ distinguishes different Floquet eigenstates, where $D$ is the dimension of the Hilbert space.  The quasienergies are defined modulo $\Omega$ as $\epsilon_\alpha \in (-\Omega/2, \Omega/2]$  and are conserved quantities. 

The time evolution operator in Floquet systems can be decomposed into two parts as~\cite{Shirley_1965,Sambe_1973}
\begin{align}
\hat{U}(t) = \hat \Phi(t)e^{-it\hat H_\mathrm{F}},
\end{align}
where $\hat{\Phi}(t) = \hat{\Phi}(t+T)$ is the time-periodic ``micromotion'' operator, and $\hat{H}_\mathrm{F}$ is the Floquet Hamiltonian. Notably, $\hat{\Phi}(0)= \hat{I}$ and $\hat{U}(nT)= e^{-i nT \hat{H}_\mathrm{F}}$. The micromotion operator provides the time evolution within a period, while the Floquet unitary operator $\hat{U}(T)$ governs the evolution at integer multiples of the period.
Using $\hat{U}_\mathrm{F} = \hat{U}(T) = e^{-iT \hat{H}_\mathrm{F}}$, or the Floquet Hamiltonian $\hat{H}_\mathrm{F} = (i/T)\log(\hat{U}_\mathrm{F})$, we can calculate the quasienergies and Floquet states for stroboscopic dynamics.

The periodic nature of the drive motivates calculating the Floquet state and its quasienergies in the frequency domain ~\cite{Shirley_1965,Sambe_1973}. The Fourier expansion of the time-periodic Hamiltonian can be written as $\hat{H}(t)=\sum_{n\in \mathbb{Z}} e^{i n\Om t} \hat{H}^{(n)}$, while the Floquet state can be expanded as $\ket{\phi_\al(t)}=\sum_{m\in \mathbb{Z}} e^{i m\Om t} \ket{\phi_\al^{(m)}}$.
The Floquet-Schr\"odinger equation, Eq.~\eqref{eq:FloqSch}, may thus be expressed as a series of equations for the various Fourier modes (labeled by $n$) of a Floquet eigenstate (labeled by $\alpha$):
\begin{equation}
\sum_{m\in \mathbb{Z}}[\hat{H}^{(n-m)}+n\Om\delta_{nm}]\ket{\phi_{\al}^{(m)}}=\ep_\al \ket{\phi_\al^{(n)}}.
\end{equation}
The above equations are a system of coupled linear equations that can also be represented in matrix form  as
\begin{equation}\label{eq:ShirleyHam}
  \begin{pmatrix}
   \dots\vdots   &  \dots    & \dots     &  \dots    & \dots   &  \dots & \dots \\
 \dots  &  \hat{H}^{(0)}-2\Om  &  \hat{H}^{(-1)} &  \hat{H}^{(-2)}    &  \dots   & \dots  & \dots \\
      \dots  &  \hat{H}^{(1)} &  \hat{H}^{(0)}-\Om  &  \hat{H}^{(-1)} &  \hat{H}^{(-2)}  & \dots & \dots \\
       \dots  & \hat{H}^{(2)}    &  \hat{H}^{(1)} &  \hat{H}^{(0)} &   \hat{H}^{(-1)}  &  \hat{H}^{(-2)}   & \dots\\
    \dots     & \dots    &  \hat{H}^{(2)}  &  \hat{H}^{(1)} &  \hat{H}^{(0)}+\Om  &  \hat{H}^{(-1)}  & \dots \\
      \dots   &   \dots  &   \dots  &  \hat{H}^{(2)}   &   \hat{H}^{(1)} &  \hat{H}^{(0)}+2\Om & \dots \\
      \dots  &  \dots   &  \dots    &   \dots   & \dots   & \dots  &   \vdots \dots
  \end{pmatrix}
  \begin{pmatrix}
 \vdots\\   
 \ket{\phi_{\al}^{(-2)}}\\
    \ket{\phi_{\al}^{(-1)}}\\
     \ket{\phi_{\al}^{(0)}}\\
      \ket{\phi_{\al}^{(1)}}\\
      \ket{\phi_{\al}^{(2)}}\\
    \vdots
  \end{pmatrix}=
  \epsilon_\alpha \begin{pmatrix}
     \vdots\\   
 \ket{\phi_{\al}^{(-2)}}\\
    \ket{\phi_{\al}^{(-1)}}\\
     \ket{\phi_{\al}^{(0)}}\\
      \ket{\phi_{\al}^{(1)}}\\
      \ket{\phi_{\al}^{(2)}}\\
    \vdots
  \end{pmatrix}.
\end{equation}
Thus, the effective time-independent extended Floquet Hamiltonian is an infinite-dimensional matrix. It is worth emphasizing that the elements of the extended Floquet Hamiltonian are Fourier modes of the time-periodic Hamiltonian and the elements of the extended Floquet state (column vector) are Fourier modes of the Floquet state $\ket{\phi_{\alpha}(t)}$.
 
Upon solving Eq.~(\ref{eq:ShirleyHam}), we obtain the quasienergies $\{ \epsilon_{\alpha}\pm n\Omega\}_{n=0}^{\infty}$ and associated Floquet eigenstates. We note that the quasienergies are unbounded in the frequency domain, and the different values of $n$ correspond to different Floquet zones. The zone with $n=0$ is termed the \textit{central Floquet zone}, and the quasienergies of this zone are identical to the quasienergies calculated directly using $\hat{U}(T)$ or $\hat{H}_\mathrm{F}$.

\begin{figure}[t]
    \centering
    \includegraphics[width=0.9\linewidth]{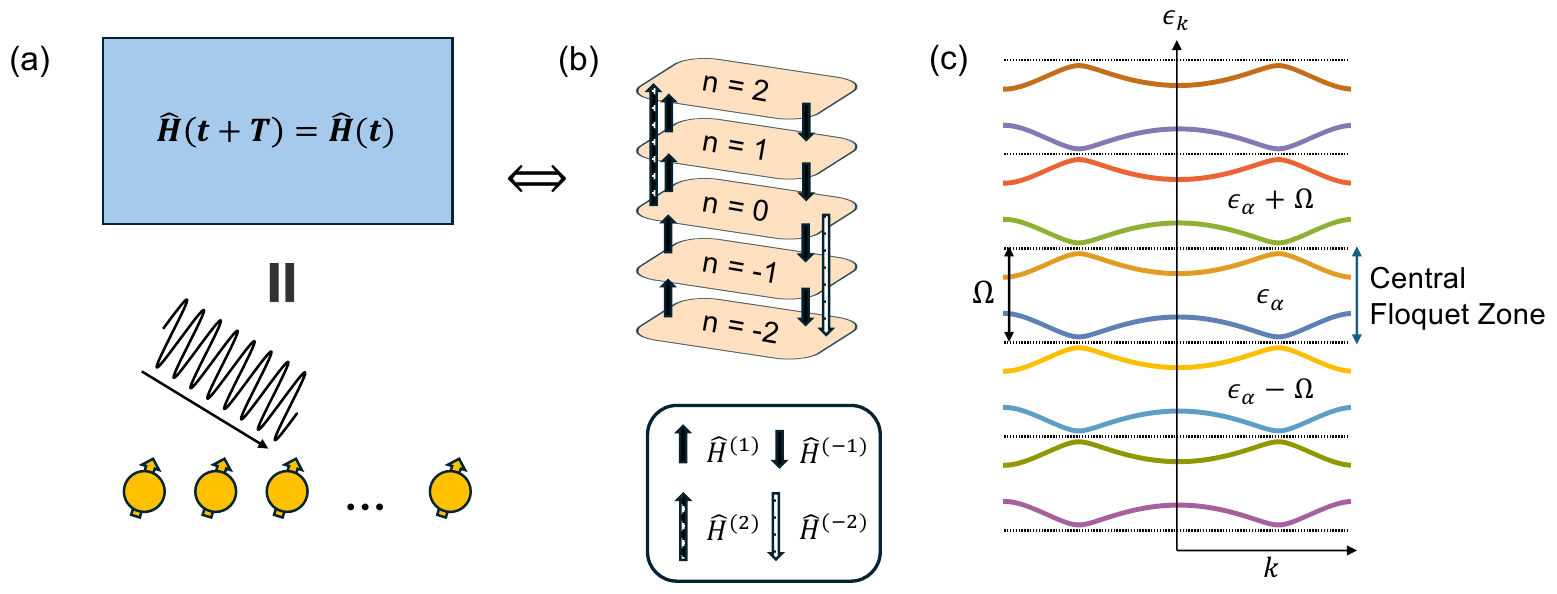}
    \caption{Schematic diagrams for a Floquet system: (a) A periodically driven system with a time-periodic Hamiltonian  $\hat H(t + T) = \hat H(t)$; (b) representation of the Fourier modes of the Floquet state (indicated by planes) with indices $n \in \mathbb{Z}$ in the frequency domain, where the Hamiltonian's Fourier components $\hat H^{(n)}$  connect different Fourier planes; and (c) quasienergies across various Floquet zones for a driven two-band model (one qubit with coefficients dependent on parameter $k$).}
    \label{Fig:Schematic}
\end{figure}
In practice, we truncate the extended Floquet Hamiltonian, retaining a finite number of Floquet zones (or Fourier indices of a Floquet state) from $- N_\mathrm{c}$ to  $N_\mathrm{c}$. An equal number of positive ($N_\mathrm{c}\geq n>0$) and negative ($-N_\mathrm{c}\leq n<0$) Floquet zones are necessary to ensure that a general $\hat{H}(t)$ is Hermitian even after truncation. Additionally, asymmetric truncation (i.e., including an unequal number of positive and negative Floquet zones) introduces more error compared to symmetric truncation. Qualitatively, it is known that the physics at higher frequencies and lower driving amplitudes can be reliably captured using fewer Floquet zones, and vice versa. We will discuss more about truncation in Sec.~\ref{subsec:Auxi_Algo_Pauli}.
For a $D\times D$-dimensional $\hat H(t)$, and a cut-off at the $\pm N_\mathrm{c}$-th Fourier index, we obtain an extended Floquet Hamiltonian of dimensions $D(2N_\mathrm{c} +1)\times D(2N_\mathrm{c} +1)$.

We can formalize the structure of the extended Floquet Hamiltonian and its corresponding Hilbert space (for more details, see Refs.~\cite{Sambe_1973,Rodriguez-Vega_2018}). We consider a physical Hilbert space $(\mathcal{H})$ and an auxiliary Hilbert space $(\mathcal{A})$ that represents the Fourier indices (Floquet zones), such that the tensor product of the two forms a Floquet-Hilbert space $\mathcal{F}=\mathcal{A}\otimes \mathcal{H}$.
The Schr\"odinger equation in the Floquet-Hilbert space can be written as
 \begin{equation}
 \hhat{H}_\mathrm{F}\kket{\phi_{\alpha n}} = (\epsilon_{\alpha}+n\Omega)\kket{\phi_{\alpha n}},
 \end{equation}
 where
 \begin{align}\label{Eq_Floq_Ham_Ext}
 \hhat{H}_\mathrm{F} &= I\otimes \hat{H}^{(0)}+\sum_{k\in \mathbb{Z}}k\Omega\ket{k}\bra{k}\otimes \hat{I} + \sum_{m\in \mathbb{F}} \sum_{ k\in \mathbb{Z}}\ket{k+m}\bra{k}\otimes \hat{H}^{(m)}, \\
\kket{\phi_{\alpha n}} &= \sum_{l\in\mathbb{Z}}\ket{l+n}\ket{\phi^{(l)}_\alpha}.
 \end{align}
Here, $\ket{l}$ is a basis state of the auxiliary space labeled by the Fourier mode $l$, and $\ket{p}\bra{q}$ represents a matrix $M$ in the auxiliary space in which $M_{p,q} = 1$, while all other elements are zero. Also, $\hat H^{(n)}$ is a Fourier component of $\hat H(t)$ and $\mathbb{F}$ denotes the set of nonzero indices of the Fourier modes corresponding to $\hat{H}(t)$. We note that $\alpha$ denotes a solution within a single Floquet zone, while $n$ denotes different Floquet zones. It is worth emphasizing the distinction between Floquet zones and the Fourier basis. We write the extended Floquet Hamiltonian in the Fourier basis. However, upon block-diagonalizing the extended Floquet Hamiltonian into blocks whose dimension equals that of the physical Hilbert space, we obtain distinct Floquet zones. In each Floquet zone, we obtain the same number of eigenvalues, as shown in Fig.~\ref{Fig:Schematic}(c) for a one-qubit driven system. We elaborate further on this subtle distinction between the Fourier basis and Floquet zones in Appendix~\ref{Appen:Block_Extd_Floq_Ham}.

\subsubsection{Simple example of a periodically driven system}
In this subsection, we consider an example of a one-qubit driven system to give an explicit expression for the extended Floquet Hamiltonian and discuss the extended Floquet state. The time-periodic Hamiltonian we consider is
\begin{equation}
     \hat H(t) = d_1 \hat{\sigma}_x + 2d_2 \hat{\sigma}_y \cos(\Omega t) + 2d_3 \hat{\sigma}_z \sin(2 \Omega t),
\end{equation}
where $d_1, d_2, d_3$ are scalars, $\Omega = 2\pi / T$, and $\hat{\sigma}_x, \hat{\sigma}_y, \hat{\sigma}_z$ are Pauli matrices.

The nonzero Fourier modes of the Hamiltonian are
\[
\hat H^{(0)} = d_1 \hat \sigma_x, \quad \hat H^{(1)} = \hat H^{(-1)} = d_2 \hat \sigma_y, \quad \hat H^{(2)} = -\hat H^{(-2)} = -i d_3 \hat \sigma_z.
\]
For truncation at five Floquet zones (five-dimensional auxiliary space, $N_\mathrm{c}=2$), Fig.~\ref{Fig:Schematic}(b) schematically shows how the different Fourier modes of the Hamiltonian connect different Fourier modes of the state, or equivalently, different auxiliary states. The extended Floquet Hamiltonian can be written as
\begin{align}
 \hhat{H}_\mathrm{F} =
 \begin{pmatrix}
       d_1 \hat{\sigma}_x - 2\Omega  &  d_2 \hat{\sigma}_y & i d_3 \hat{\sigma}_z & 0 & 0 \\
       d_2 \hat{\sigma}_y &  d_1 \hat{\sigma}_x - \Omega & d_2 \hat{\sigma}_y & i d_3 \hat{\sigma}_z & 0 \\
       -i d_3 \hat{\sigma}_z & d_2 \hat{\sigma}_y & d_1 \hat{\sigma}_x & d_2 \hat{\sigma}_y & i d_3 \hat{\sigma}_z \\
       0 & -i d_3 \hat{\sigma}_z & d_2 \hat{\sigma}_y & d_1 \hat{\sigma}_x + \Omega & d_2 \hat{\sigma}_y \\
       0 & 0 & -i d_3 \hat{\sigma}_z & d_2 \hat{\sigma}_y & d_1 \hat{\sigma}_x + 2\Omega
  \end{pmatrix}.
\end{align}
This matrix can be decomposed into several terms using auxiliary matrices:
\begin{align}
\hhat{H}_\mathrm{F} &=
\begin{pmatrix}
       1 & 0 & 0 & 0 & 0 \\
       0 & 1 & 0 & 0 & 0 \\
       0 & 0 & 1 & 0 & 0 \\
       0 & 0 & 0 & 1 & 0 \\
       0 & 0 & 0 & 0 & 1
\end{pmatrix} \otimes d_1 \hat{\sigma}_x
+  
\begin{pmatrix}
       -2 & 0 & 0 & 0 & 0 \\
       0 & -1 & 0 & 0 & 0 \\
       0 & 0 & 0 & 0 & 0 \\
       0 & 0 & 0 & 1 & 0 \\
       0 & 0 & 0 & 0 & 2
\end{pmatrix} \otimes \Omega \hat{I}
+
\begin{pmatrix}
       0 & 1 & 0 & 0 & 0 \\
       1 & 0 & 1 & 0 & 0 \\
       0 & 1 & 0 & 1 & 0 \\
       0 & 0 & 1 & 0 & 1 \\
       0 & 0 & 0 & 1 & 0
\end{pmatrix} \otimes d_2 \hat{\sigma}_y
\nonumber\\
&\hspace{3in}+\begin{pmatrix}
       0 & 0 & i & 0 & 0 \\
       0 & 0 & 0 & i & 0 \\
       -i & 0 & 0 & 0 & i \\
       0 & -i & 0 & 0 & 0 \\
       0 & 0 & -i & 0 & 0
\end{pmatrix} \otimes d_3 \hat{\sigma}_z.
\end{align}

After diagonalizing \(\hhat{H}_\mathrm{F}\), we obtain 10 eigenvalues, corresponding to two quasienergies per Floquet zone. Once the quasienergies in the central Floquet zone are known, neglecting the error due to truncation, the quasienergies in any other Floquet zone can be obtained by shifting the central zone quasienergies by an integer multiple of $\Omega$. Since the central Floquet zone contains all the necessary information and provides the most accurate results, it is sufficient to focus on this zone. The extended Floquet states in the central Floquet zone (\(n=0\)) are given by
\[
(\ket{\phi_\alpha^{(-2)}}, \ket{\phi_\alpha^{(-1)}}, \ket{\phi_\alpha^{(0)}}, \ket{\phi_\alpha^{(1)}}, \ket{\phi_\alpha^{(2)}})^\mathrm{T}, \text{ where } \alpha\in\{0,1\}.
\]
 We can express the Floquet states as
\begin{align}
\ket{\phi_\alpha(t)} = \sum_{l=-2}^{2} e^{il\Omega t} \ket{\phi^{(l)}_\alpha}.
\end{align}

We note that $\ket{\phi_\alpha(0)}$ is an eigenstate of $\hat{H}_\mathrm{F}$ with eigenvalue $\epsilon_\alpha$. To fully describe the time-periodic Floquet state, we also need to know the micromotion operator $\hat{\Phi}(t)$, where $\ket{\phi_\alpha(t)} = \hat{\Phi}(t) \ket{\phi_\alpha(0)}$.

When $d_1$, $d_2$, and $d_3$ depend on a parameter or set of parameters $\vec{k}$, two quasienergy bands are obtained in each Floquet zone, as shown schematically in Fig.~\ref{Fig:Schematic}(c). In the central Floquet zone, the quasienergies are $\epsilon_\alpha(\vec{k})$, and the corresponding extended Floquet state is $(\ket{\phi_\alpha^{(-2)}(\vec{k})}, \ket{\phi_\alpha^{(-1)}(\vec{k})}, \ket{\phi_\alpha^{(0)}(\vec{k})}, \ket{\phi_\alpha^{(1)}(\vec{k})}, \ket{\phi_\alpha^{(2)}(\vec{k})})^\mathrm{T}$, where $\alpha \in \{0,1\}$. In solid-state systems, $\vec{k}$ can be thought of as the crystal momentum.

\begin{figure}[t]
    \centering
    \includegraphics[width=0.9\linewidth]{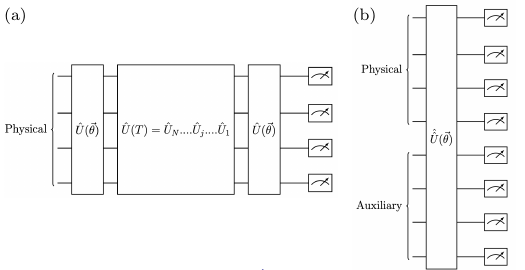}
    \caption{Schematic diagram of (a) the deeper quantum circuit needed to realize the Floquet unitary ($\hat U_\mathrm{F}=\hat U(T)$) versus (b) the shallower and wider circuit needed for preparing eigenstates of the extended Floquet Hamiltonian.}
    \label{UT_HF_circuits}
\end{figure}

We emphasize here that there are two distinct approaches to studying quasienergies and Floquet states in periodically driven systems. These are: (i) using the Floquet unitary ($\hat{U}_\mathrm{F}$), and (ii) preparing eigenstates of the extended Floquet Hamiltonian. In the case of a continuous periodic drive, obtaining  $\hat{U}_\mathrm{F}$ through Trotterization requires a much deeper circuit~\cite{Smith2019}. In Fig.~\ref{UT_HF_circuits}, we schematically show the deeper circuit required to implement $\hat{U}_\mathrm{F}$, compared to the shallower and wider circuit that results from applying the extended Floquet Hamiltonian framework. Therefore, the extended Floquet Hamiltonian framework is likely to be more efficient for studying quantum many-body systems subject to continuous driving on NISQ devices.

\subsection{ADAPT-VQE} \label{sec:ADAPT_basics}
The ADAPT-VQE algorithm~\cite{Grimsley_NatCom_2019} employs a greedy strategy to iteratively construct a variational ansatz using operators from a fixed, user-defined operator pool, starting from an initial state $|\psi_{\mathrm{ref}}\rangle$. Let $\mathcal{O} = \{ \hat O^{(1)}, \hat O^{(2)}, \dots \hat O^{(N)}\}$ be the pool of operators.  Then, defining $\vect{\theta}^{(k)} = (\theta_1,\dots,\theta_k)$ to be the variational parameters in the ansatz at iteration $k$, the ansatz $|\psi_{k}(\vect{\theta}^{(k)})\rangle$ is grown by appending operator $\hat O_{k+1}\in\mathcal{O}$ to it, with a corresponding variational coefficient $\theta_{k+1}$, to generate the new ansatz
\begin{equation}\label{adapt-vqe-eqn}
    |\psi_{k+1}(\vect{\theta}^{(k+1)})\rangle = e^{-i\theta_{k+1 }\hat O_{k+1}}|\psi_{k}(\vect{\theta}^{(k)})\rangle.
\end{equation}

In Eq.~\eqref{adapt-vqe-eqn}, the operator $\hat O_{k+1}$ that is appended to the previous ansatz is chosen by measuring the energy gradient of each pool operator $\left| \left. \partial \langle \hat H \rangle / \partial \theta_{k+1}\right|_{\theta_{k+1}=0} \right|$ at the state $|\psi_{k}(\vect{\theta}^{(k)})\rangle$ and selecting the operator that has the largest gradient. The energy gradient of a pool operator can be expressed as
\begin{equation}\label{adapt-gradient}
    \left| \left. \partial \langle \hat H \rangle / \partial \theta_{k+1}\right|_{\theta_{k+1}=0} \right| =  \left| \langle \psi_{k}(\vect{\theta}^{(k)}) | \left[ \hat O_{k+1}, \hat H \right] | \psi_{k}(\vect{\theta}^{(k)}) \rangle \right|,
\end{equation}
where the right-hand side of Eq.~\eqref{adapt-gradient} can be efficiently measured on a quantum processor. A convergence check follows the measurement of the pool operator gradients: If the gradient norm of the pool operators is smaller than a threshold $\varepsilon$, the calculation is terminated and declared converged; if not, the ansatz is grown as in Eq.~\eqref{adapt-vqe-eqn}. Finally, a VQE optimization of all variational parameters is performed, in which the parameters are initialized as $\vect{\theta}_{k+1}: (\theta_1 = \theta_1^*, \dots, \theta_k=\theta_k^*, \theta_{k+1} = 0)$, i.e., the parameters $(\theta_1,\dots,\theta_k)$ take their previously optimized values, and the newly added parameter $\theta_{k+1}$ is initialized to $0$. The steps of growing the ansatz and re-optimizing the coefficients are repeated until the convergence criterion is met. 

The ADAPT-VQE algorithm was demonstrated to reach arbitrarily accurate energies and yield shallow ansätze in simulations of small molecules~\cite{Grimsley_NatCom_2019,Tang_QubitADAPT_2021,ramôa_2024_CEOPool}. Recently, there have been improvements, including adding operators in batches to further improve circuit compactness, as in the TETRIS-ADAPT-VQE algorithm~\cite{Anastasiou_Tetris2024}. It has also been extended to a diverse range of problems beyond chemistry, including ground-state preparation for strongly-correlated systems in high-energy~\cite{Farrell2024,gustafson2024surrogateconstructedscalablecircuits} and condensed matter~\cite{Gyawali_2022_FermiHubbard,Dyke_Shirali_2024_tiling} physics, thermal state preparation~\cite{warren2022adaptivevariationalalgorithmsquantum}, as well as the calculations of excited states~\cite{yordanov2021molecular}, time-evolution~\cite{Gomes2021} and dynamics~\cite{Yao2021}. Its versatility makes it especially suitable for the task of generating shallow circuits that prepare extended Floquet states.

\section{Floquet-ADAPT-VQE algorithm}\label{sec:Floq_ADAPT_Algo}

We begin this section by developing a method to decompose the auxiliary matrices into a Pauli basis to facilitate mapping the extended Floquet Hamiltonian onto a quantum processor. To formulate Flqouet-ADAPT-VQE algorithm, we need to define a cost function, find a suitable initial state, and choose pool operators. This algorithm prepares extended Floquet states on a quantum processor and obtains the full spectrum of quasienergies. Finally, we propose quantum circuit to calculate the expectation value of an operator after an extended Floquet state has been prepared. 

\subsection{Decomposition of auxiliary matrices into  Pauli strings}\label{subsec:Auxi_Algo_Pauli}
We consider a time-periodic Hamiltonian in a $D$-dimensional Hilbert space. For convenience we choose $D=2^L$. In Eq.~(\ref{Eq_Floq_Ham_Ext}), the extended Floquet Hamiltonian contains an odd number $(2N_\mathrm{c} +1)$ of Floquet zones and so has an odd-dimensional auxiliary space. However, in order to implement our Hamiltonian on qubit-based quantum hardware, we require the dimension to be a power of 2. For this purpose, we choose the number $n_a$ of auxiliary qubits and the cut-off $N_\mathrm{c}$ such that $2(N_\mathrm{c} +1) = 2^{n_a}$, and arrange the extended Hamiltonian into two blocks. The larger block contains $2N_\mathrm{c} +1$ Floquet zones and the smaller contains one Floquet zone (see Appendix \ref{Appen:Block_Extd_Floq_Ham}). We can write the extended Floquet Hamiltonian as   
\begin{equation} \label{Eq:HF_Ext}
    \hhat{H}_\mathrm{F}=  I \otimes \hat H^{(0)} + \Omega A^{\mathrm{d}}\otimes  \hat I + \sum_{r\in \mathbb{F}}(A^{(r)}-A_{\mathrm{asy}}^{(r)})\otimes \hat H^{(r)} ,
\end{equation}
where $I$ ($\hat I$) is the identity matrix in the auxiliary (physical) Hilbert  space, and $ A^{\mathrm{d}}=\sum_{n} n\ket{n}\bra{n}$, $ A^{(r)}= \sum_{n}\ket{n+r}\bra{n}$ are respectively diagonal and off-diagonal matrices in the auxiliary space, where the summations are over all the states in the auxiliary space for which $-N_\mathrm{c}\le n\le N_\mathrm{c}+1$ and $-N_\mathrm{c}\le n+r\le N_\mathrm{c}+1$. In the summation in Eq.~\eqref{Eq:HF_Ext}, $\mathbb{F}$ denotes the set of indices labeling the nonzero Fourier modes  of the time-periodic Hamiltonian. The final term in Eq.~(\ref{Eq:HF_Ext}) involving $A_{\mathrm{asy}}^{(r)}$ is subtracted to ensure that the extended Floquet Hamiltonian is block-diagonalized and that the larger blocks remain symmetric across the Floquet zones. Here, $\hat H^{(r)}$ is a Fourier mode of the time-periodic Hamiltonian $\hat H(t)$ and belongs to the physical Hilbert space ($\mathcal{H}$).  

We decompose the auxiliary diagonal matrix $A^{\mathrm{d}}$ and off-diagonal matrices $A^{(r\neq 0)}$ in terms of Pauli strings (matrices). Naively, we could use a trace-based method:
\begin{align}\label{Eq.:ADAPT_algo}
   A^{(r)}=  \sum_{j=1}^{4^{n_a}} d_j{P_j}, \ \ \ d_j= \frac{1}{2^{n_a}}\Tr[A^{(r)} P_j],
\end{align}
where $P_j\in \{ I,X,Y,Z\}^{\otimes n_a}$ and $\{d_j\}_{j=1}^{4^{n_a}}$ are complex coefficients. However, this approach is not efficient as it requires calculating $4^{n_a}$ coefficients, while the structure of the auxiliary matrices allows them to be decomposed into a much smaller subset of Pauli strings in which most of the $4^{n_a}$ coefficients $d_j$ are in fact zero. We now present the explicit form of these decompositions. 

The diagonal auxiliary matrix $A^{\mathrm{d}}$ can be decomposed into Pauli terms $P_j\in \{ I,Z\}^{\otimes n_a}$ with coefficients $\{d_j\}$. Although a generic diagonal matrix has $2^{n_a}$ such terms, the specific form of $A^{\mathrm{d}}$ in the extended Floquet Hamiltonian is comprised of only $n_a+1$ Pauli strings. In particular, we can write $A^{\mathrm{d}}= \sum_{j} d_j Z_{(j)} + \frac{1}{2} I $, where $Z_{(j)}= I_{n_a-1}\otimes\dots\otimes I_{j+1}\otimes Z_{j}\otimes I_{j-1}\dots\otimes I_0$. Here, the coefficients are given by $d_j= -2^{j-1}$.
For the off-diagonal matrices $A^{(r\neq 0)}$, we have 
\begin{equation}\label{eq:Am1}
A^{(-1)}=\sum_{n=-N_\mathrm{c}}^{N_\mathrm{c}}\ket{n}\bra{n+1}=\sum_{k=0}^{n_a-1}I^{\otimes k}\otimes P^+\otimes (P^-)^{\otimes n_a-k-1},
\end{equation}
\begin{equation}\label{eq:A1}
A^{(1)}=A^{(-1)\dagger}=\sum_{n=-N_\mathrm{c}}^{N_\mathrm{c}}\ket{n+1}\bra{n}=\sum_{k=0}^{n_a-1}I^{\otimes k}\otimes P^-\otimes (P^+)^{\otimes n_a-k-1},
\end{equation}
\begin{equation}\label{eq:Ar}
A^{(|r|)}=[A^{(1)}]^{|r|}, \ \ \  A^{(-|r|)}=A^{(|r|)\dagger}=[A^{(-1)}]^{|r|},
\end{equation}
where $P^{\pm}=(X\pm iY)/2$. The number of Pauli strings in each of these operators is $O(2^{n_a})$ (see Appendix \ref{Appen:Cal_Paui_Off}). 

For the auxiliary matrices that we use to correct for truncation asymmetries, we find
\begin{eqnarray}\label{eq:Aasy}
    A_{\mathrm{asy}}^{(r>0)}&=&\ket{N_\mathrm{c}+1}\bra{N_\mathrm{c}+1-r}=(P^-)^{x_{n_a-1}}(Z^-)^{1-x_{n_a-1}}\otimes\ldots\otimes(P^-)^{x_0}(Z^-)^{1-x_0},\\
    A_{\mathrm{asy}}^{(r<0)}&=&\ket{N_\mathrm{c}+1-|r|}\bra{N_\mathrm{c}+1}=(P^+)^{x_{n_a-1}}(Z^-)^{1-x_{n_a-1}}\otimes\ldots\otimes(P^+)^{x_0}(Z^-)^{1-x_0},
\end{eqnarray}
where
\begin{equation}
    x_i=\left\lfloor{\frac{|r|}{2^i}}\right\rfloor\mod 2,
\end{equation}
and $Z^{\pm}=(I\pm Z)/2$. In other words, we first express $|r|$ as an $n_a$-bit string (say, $r_{\text{bit}}$). Then we replace 0 with $Z^{-}$ and 1 with $P^{-}$  (for $A_{\mathrm{asy}}^{(r>0)}$) or  $P^{+}$  (for $A_{\mathrm{asy}}^{(r<0)}$) in $r_{\text{bit}}$. For example, if $n_a = 5$ and $r = 3$, we have $r_{bit} = 00011$ and so $A_{\mathrm{asy}}^{(3)}=Z^{-}\otimes Z^{-}\otimes Z^{-}\otimes P^{-}\otimes P^{-}$, $A_{\mathrm{asy}}^{(-3)}=Z^{-}\otimes Z^{-}\otimes Z^{-}\otimes P^{+}\otimes P^{+} $.
Thus, each asymmetric auxiliary matrix can be decomposed into $2^{n_a}$ Pauli strings, and we find that the number of Pauli strings needed to express auxiliary matrices is $O(2^{n_a})$, which is much less than the total number of Pauli strings $(4^{n_a})$. 

For any quantum simulation, converting a Hamiltonian into a sum of Pauli strings is the first and necessary step toward its implementation on a qubit-based quantum processor. Therefore, we would like to emphasize that the Pauli decomposition of the auxiliary matrices and the extended Floquet Hamiltonian serves as a basic input for various algorithms in both near-term and fault-tolerant eras. For example, this decomposition can also be used for simulating quantum dynamics via the time-evolution unitary of the extended Floquet Hamiltonian in the Floquet-Hilbert space \cite{Levante_Floq_Dynamics,Kaoru_Optimal_Floquet_2023}.

In order to efficiently simulate an $L$-qubit time-independent Hamiltonian on a quantum processor, the number of Pauli strings representing the Hamiltonian should scale polynomially with $L$. For VQEs, we can  efficiently measure the energy  expectation value if the scaling of the Pauli strings is polynomial~\cite{Peruzzo_2014_VQE}.  However, in the case of an extended Floquet Hamiltonian, the number of Pauli strings required to represent auxiliary matrices scales as $2^{n_a} = 2(N_{\mathrm{c}} + 1)$. This implies that, even if the Hamiltonian of the physical space contains a polynomial number of Pauli strings, the total number of Pauli strings in the extended Floquet Hamiltonian will scale as $O\big(2^{n_a} \times \text{poly}(L)\big)$. Therefore, to efficiently simulate the extended Floquet Hamiltonian on a quantum computer, the number of Pauli strings in the auxiliary matrices must scale polynomially with the number of physical qubits $L$, i.e., $2^{n_a} = \text{poly}(L)$. Fortunately, for scenarios involving only a few Fourier modes \cite{rudner2020FloquetEngineersHandbook}, accurate truncation of the extended Floquet Hamiltonian is achievable by taking $n_a= C\log_2(LJ_{\textrm{max}}/\Omega)$ where $J_{\textrm{max}}$ is the maximum value of local energy parameters and $C$ is a constant. In Appendix~\ref{Appen:truncation}, we discuss the truncation in detail and show that, for a driven $L$-qubit system with $O(L^K)$ Pauli strings, we have $C=K$. Thus, higher frequencies and lower values of $J_{\text{max}}$ require fewer auxiliary qubits, and vice versa. A more rigorous derivation of the auxiliary-qubits ($n_a$) requirement is provided in Ref.~\cite{Mizuta_FloquetQPE_2025}, where $n_a = \Theta\left(\log(\alpha T) + \log\log(1/\varepsilon)\right)$. Here, $\varepsilon$ denotes the  error tolerance in quasienergy due to truncation, $T$ is the drive period  and $\alpha = \max\{||H^{(m)}||\}$, which is proportional to $LJ_{\text{max}}$.

\subsection{Floquet-ADAPT-VQE algorithm for preparing Floquet states} \label{sec:Prep_Floq_state}
We now describe our Floquet-ADAPT-VQE algorithm for preparing Floquet states and measuring quasienergies on a quantum computer. We must first identify a suitable cost function. The expectation value of the extended Floquet Hamiltonian $\hhat{H}_\mathrm{F}$ might seem to be a natural choice. However, a variational algorithm based on such a cost function would target Floquet states corresponding to the lowest Floquet zone that remains when we truncate the auxiliary space, which is where the truncation errors are largest. Since each Floquet zone contains the same quasienergy spectrum up to a constant shift by an integer multiple of $\Omega$, it is instead preferable to target the central Floquet zone, where the truncation errors are minimal. This motivates us to consider a cost function that involves the square of the extended Floquet Hamiltonian shifted by a constant $\lambda$:
\begin{equation}\label{Eq:Cost_function}
    C_\lambda(\vec{\theta})= \bbra{\psi(\vec{\theta})}(\hhat{H}_\mathrm{F}-\lambda \hhat{I})^2 \kket{\psi(\vec{\theta})}.
\end{equation}

The shift parameter $\lambda$ can be chosen to target a quasienergy close to $\lambda$, in a manner analogous to similar methods used for solving static problems variationally~\cite{Wang_1994}. We choose  $-\Omega/2 < \lambda < \Omega/2$ to find quasienergies in the central Floquet zone. Here, $\vec{\theta}$ are the variational parameters at each ADAPT iteration as discussed in Sec.~\ref{sec:ADAPT_basics}, and $\kket{\psi(\vec{\theta})}$ is the normalized extended Floquet state.
The cost function is strictly nonnegative and attains its minimal value when $\kket{\psi(\vec{\theta})}$ is the Floquet state whose quasienergy is closest to $\lambda$. By minimizing the cost function using the ADAPT-VQE algorithm, we thus expect to obtain the quasienergy closest to $\lambda$ and its corresponding state. We denote $\vec{\theta}_0= \text{argmin}\big(C_\lambda(\vec{\theta}\big)\big)$ at the final ADAPT iteration. We can test whether the state is the actual Floquet state by taking  the expectation value of $\hhat{H}_\mathrm{F}$. If we find that $\bbra{\psi(\vec{\theta}_0)}\hhat{H}_\mathrm{F}\kket{\psi(\vec{\theta}_0)}\in \{\pm\sqrt{C_\lambda(\vec{\theta}_0)} + \lambda\}$ then we have prepared the extended Floquet state and the quasienergy is $\bbra{\psi(\vec{\theta}_0)}\hhat{H}_\mathrm{F}\kket{\psi(\vec{\theta}_0)}$.

Now, we set $\lambda = 0$ to define the cost function as $C_0(\vec{\theta})$.  The eigenstates of $\hhat{H}_\mathrm{F}^2$ are also eigenstates of $\hhat{H}_\mathrm{F}$, except for cases where quasienergies appear in pairs of $\pm \epsilon$. This pairing occurs when $\hhat{H}_\mathrm{F}$ exhibits chiral symmetry or particle-hole symmetry, where there exists an operator $\hhat{\Gamma}$ satisfying $\{\hhat{H}_\mathrm{F}, \hhat{\Gamma}\} = \hhat{H}_\mathrm{F}\hhat{\Gamma} + \hhat{\Gamma}\hhat{H}_\mathrm{F} = 0$. After minimizing the cost function, we obtain a quantum state and calculate the expectation value of the extended Floquet Hamiltonian $\hhat{H}_\mathrm{F}$. If the expectation value is $\pm \sqrt{C_0(\vec{\theta}_0)}$, the extended Floquet state has been prepared. Otherwise, we have a superposition of extended Floquet states given by $c_1\kket{\phi(\vec{\theta}_0)} + c_2 \hhat{\Gamma}\kket{\phi (\vec{\theta}_0)}$, with $|c_1|^2 + |c_2|^2 = 1$ and $\kket{\phi}$ is an eigenvector with eigenvalue $\epsilon$.


Next, for $\lambda\neq 0 $ the cost function targets a specific quasienergy and Floquet state. Here, even if $\hhat{H}_\mathrm{F}$ has chiral symmetry or particle-hole symmetry, $\hhat{H}_\mathrm{F}-\lambda \hhat{I}$ does not. By choosing a very small value of $\lambda$, we can prepare a Floquet state even in the presence of chiral symmetry or particle-hole symmetry. Moreover, by treating $\lambda$ as a tunable parameter and discretizing it, we can determine the full quasienergy spectrum.  To capture all extended Floquet states within the central Floquet zone, we may need to employ a very fine mesh for $\lambda$. However, the key advantage of this approach compared to other methods, like the one used in Ref.~\cite{Higgott2019variationalquantum,Fauseweh2023} is that it does not require including additional penalty terms in the cost function to project out previously prepared eigenstates, which significantly increase the measurement costs. 

The choice of the initial state significantly affects the convergence of VQE. We focus on targeting quasienergy states in the central Floquet zone ($0 < \epsilon^2 < \Omega^2/4$) since the truncated extended Floquet Hamiltonian provides the most accurate Floquet states within this zone. However, the full spectrum of squared quasienergies extends from zero to $(N_\mathrm{c} + 1/2)^2 \Omega^2$, where $N_\mathrm{c}$ represents the number of truncated positive Floquet zones. A randomly chosen state is more likely to be further away from the central Floquet zone as we increase $N_\mathrm{c}$ or $n_a$. This adversely affects the convergence of a VQE. Therefore, it is crucial to identify an auxiliary state that brings the squared Floquet energy ($C_0(\vec{\theta})$) closer to the central Floquet zone. We choose a state which is an eigenstate of $A^{\mathrm{d}}$ with zero eigenvalue: 
\begin{align}\label{Eq:init_state}
    \kket{\psi_\mathrm{in}} = \ket{0}\otimes\ket{\phi_\mathrm{p}},
\end{align}
where $\ket{0}$ is the auxiliary state associated with the central Floquet zone that satisfies $A^{\mathrm{d}}\ket{0}= 0$, and where $\ket{\phi_\mathrm{p}}$ is an initial state of the physical system.
In the standard basis of $\mathbb{R}^{2^{n_a}}$ in which $\ket{-N_\mathrm{c}}$ is represented as $(1,0,...,0)^\mathrm{T}$, this auxiliary state is $\ket{0}\equiv (0,...0,1,0,...,0)^\mathrm{T}$, where there is a single $1$ in the $2^{n_a -1}$th component. In the computational basis of the $n_a$-qubit auxiliary space in which $\ket{-N_\mathrm{c}}\equiv\ket{\uparrow_{n_a-1}\uparrow_{n_a-2}...\uparrow_1\uparrow_0}$, $\ket{0}\equiv\ket{\uparrow_{n_a-1}\downarrow_{n_a-2}...\downarrow_1\downarrow_0}$, where  $\ket{\uparrow}\equiv(1 \ 0)^{\mathrm{T}}, \  \ket{\downarrow}\equiv(0 \ 1)^{\mathrm{T}}$. The cost function value for this choice of auxiliary initial state  is independent of the number of auxiliary qubits $n_a$, as well as the frequency $\Omega$ (see Appendix~\ref{Appen:initial_state} for a proof), thereby keeping the energy of the initial state close to or within the central Floquet zone for arbitrary truncation and driving frequency. Further improvements to the initial state can be made by optimizing the choice of $\ket{\phi_\mathrm{p}}$ based on the specific physical system under consideration. 

To run Floquet-ADAPT-VQE, we need to define an operator pool that aids in creating the ansatz in a problem-informed manner. The simplest (though not necessarily the most efficient) choice for the operator pool is the standard pool of Pauli strings used for a static system with $L+n_a$ qubits. However, for systems that require more auxiliary qubits, this might lead to an unreasonable number of gradient measurements. To prevent this, we construct a mixed Floquet pool by using different operator pools for the auxiliary and physical parts of the Floquet-Hilbert space. For example, we might select the full Pauli pool for the auxiliary part and a pool of two-local operators for the physical part in the case of a spin-lattice system. Depending on the type of physical system, various operator pools can be chosen based on existing works on static Hamiltonians \cite{Tang_QubitADAPT_2021,Yordanov_2021_QEBPool,Shkolnikov_2023_symmetryPool,Dyke_Shirali_2024_tiling,ramôa_2024_CEOPool}. We summarize the full Floquet-ADAPT-VQE algorithm in Algorithm~1.

\begin{algorithm}[H]\label{algo:Floq_ADAPT}
  \caption{Floquet-ADAPT-VQE}\label{algo:Floq_ADAPT1}
  \begin{algorithmic}

    \REQUIRE Fourier modes of the time-periodic Hamiltonian
    \ENSURE Quasienergies and Floquet state in Floquet-Hilbert space prepared on a quantum processor

    \STATE 1. Construct all the auxiliary operators $A^{(r)}$ and construct $\hhat{H}_\mathrm{F}^2$ in the Floquet-Hilbert space
    
    \STATE 2. Choose an initial state as $\kket{\psi_\mathrm{in}} = \ket{0} \otimes\ket{\phi_\mathrm{p}}$
    
    \STATE 3. Select Floquet operator pools (a mix of auxiliary and physical pools) in Floquet-Hilbert space
    
    \STATE 4. Apply the ADAPT-VQE subroutine on $\hhat{H}_\mathrm{F}^2$ to obtain the Floquet state and squared quasienergies closest to zero
    
    \STATE 5. To obtain the full quasienergy spectrum within the central Floquet zone, discretize $\lambda$ finely and apply the ADAPT-VQE subroutine on $(\hhat{H}_\mathrm{F} - \lambda \hhat{I})^2$. Adjust the initial state as needed to capture the full spectrum
    
    \STATE 6. To find the exact quasienergy, compute the expectation value of $\hhat{H}_\mathrm{F}$ in the Floquet state

  \end{algorithmic}
\end{algorithm}

We now briefly discuss the quantum resources---auxiliary qubits and measurement cost---required for implementing Floquet-ADAPT-VQE. We assume a driven $L$-qubit Hamiltonian containing $O(L^K)$ Pauli strings, where the maximum local interaction strength, or equivalently the maximum coefficient of a Pauli string ($J_{\textrm{max}}$), and the driving frequency are fixed. As discussed in Appendix~\ref{Appen:truncation}, the number of ancilla (auxiliary) qubits scales as $O(K\log_2 L)$, so a driven $L$-qubit system requires $L + O(K\log_2 L)$ qubits to represent the extended Floquet Hamiltonian.

The measurement cost in Floquet-ADAPT-VQE consists of squared Floquet energy measurements and pool-operator-gradient measurements. For the energy measurements, the worst-case scaling is governed by the number of Pauli strings. The auxiliary part contributes $O(2^{n_a}) = O(L^K)$ Pauli strings, while the contribution from the physical space depends on the number of Pauli strings in the Fourier components of the Hamiltonian. If the Pauli terms in the time-periodic Hamiltonian scale as $O(L^K)$, then the squared extended Floquet Hamiltonian contains $O(2^{2n_a} L^{2K}) = O(L^{4K})$ Pauli strings.

The worst-case scaling of pool-gradient measurements depends on the product of the operator pool size and the number of Pauli strings in the squared extended Floquet Hamiltonian. If the pool size scales as $O(L^{P})$, then the total measurement cost scales as $O(L^{4K+P})$. This represents a worst-case estimate; in practice, the number of relevant terms can be significantly smaller due to structure and commutativity. In Sec.~IV, we explicitly discuss this for a driven XYZ model.

The above analysis is based solely on the number of Pauli terms and does not account for their grouping into commuting sets. However, for many Hamiltonians, the Pauli terms can be efficiently partitioned into commuting groups, leading to a reduced measurement-cost scaling. In Appendix~\ref{Appen:Meaurement_Scaling_of_cost_fuction}, we show that the number of commuting sets corresponding to $\hat{\hat{H}}_F^{2}$ scales as $O\left(L\log^2 L\right)$ for a time-periodic Hamiltonian with a single Fourier mode and geometrically local Pauli strings of maximum weight $w$. This implies that the measurement cost corresponding to the cost function scales as $O\left(\log^2(L)\,L\right)$. The gradient measurement cost will depend on the scaling of the pool operators.
We construct the mixed Floquet operator pool from all tensor products of weight-2 Pauli strings acting on the auxiliary space and weight-2 Pauli strings acting on the physical space. The number of Pauli strings in this pool then scales as $O\left(n_a^2 L^2\right)$. When we consider a time-periodic Hamiltonian with a single Fourier mode and $w$-weighted geometrically local Pauli strings, we obtain $O\left(n_a^4 L^3\right)= O\left(\mathrm{polylog}(L)\,L^3\right)$.

\subsection{Advantages of using Floquet-Hilbert space}\label{sec: Fixed_depth_obs_state}
The extended Floquet Hamiltonian provides a convenient way to transform a time-dependent Hamiltonian into a time-independent form, enabling the use of algorithms developed for static Hamiltonians. However, this is not the only benefit: this formalism also allows us to extract more information. In particular, the eigenstates of the extended Floquet Hamiltonian $\hhat{H}_\mathrm{F}$ contain more information than the eigenstates of the physical Floquet Hamiltonian $\hat H_F$. The latter eigenstates are also eigenstates of the physical evolution operator at discrete times, $\hat U(kT)=e^{-i\hat H_FkT}$ for integers $k$, and as such they only contain information about the time-evolved Floquet state $\ket{\phi(t)}$ at these discrete times, and not at intermediate times. In contrast, the extended Floquet state contains all Fourier components ${\ket{\phi^{(n)}}}$ of $\ket{\phi(t)}$. Consequently, in principle, the full time-dependence of $\ket{\phi(t)}$ at all times is known, including the micromotion during each Floquet period. Below, we show that the expectation value of an operator in a Floquet state can be calculated using a constant-depth circuit. We also show that $\ket{\phi(t)}$ can be prepared at arbitrary times with the help of measurements, without any additional circuit depth.

\subsubsection{Expectation value of operator in Floquet state in time}
Now we provide an algorithm to calculate the expectation value of an operator $\hat O$ evolving in time using the extended Floquet state. 
We can write the expectation value in the extended Floquet-Hilbert space as (see Appendix~\ref{App:Floq_Expt} for details)
\begin{align} \label{Eq:Expt_Expression}
   \braket{\hat O(t)}_{\phi}=\sandwich{\phi(t)}{\hat{O}}{\phi(t)} = \bbra{\phi}\hhat U^{\dagger} (t)\hhat O\hhat U (t)\kket{\phi},
\end{align}
where $\hhat U (t)=e^{iA^{\mathrm{d}}\Omega t}\otimes \hat{I}$ such that $A^\mathrm{d}$ is the diagonal auxiliary matrix and $\hhat{O}= A_o\otimes\hat{O}$, where $A_o$ is a real matrix with all elements being one [see Appendix C]. For an $n_a$-qubit auxiliary space, $e^{iA^{\mathrm{d}}\Omega t}= e^{i\Omega t/2}\otimes_{j} R_j^z(\Omega t)$, and $R_j^z(\Omega t)=e^{-i(2^{j-1}\Omega t)Z_j} $ is a single qubit $Z$-rotation operator and $j\in\{0,1,\dots,n_a-1\}$. The tensor product is ordered such that the $j=n_a-1$ operator is the left-most, and the $j=0$ operator is the right-most. For $\hhat{O}= A_o\otimes\hat{O}$, we can express $A_{o}$  in terms of all possible products of Pauli $X$ and identity $I$ as $A_{o}= (I+X)^{\otimes n_a} =\sum_{j=0}^{2^{n_a} -1} P_j$ such that $P_j\in \{I,X\}^{\otimes n_a}$. We note that all the coefficients are equal to one in this summation of Pauli operators. We can further simplify the expectation value of an operator:
\begin{align}
   \braket{\hat O(t)}_{\phi} &=\bbra{\phi}\hhat U^{\dagger} (t)\hhat O\hhat U (t)\kket{\phi} \nonumber \\
   &=\bbra{\phi} \big[\{\otimes_j R_j^z(-\Omega t)\}\otimes \hat{I}\big]\big(A_o \otimes \hat O\big) \big[\{\otimes_j R_j^z(\Omega t)\}\otimes \hat{I}\big]\kket{\phi} \nonumber\\
   &=\bbra{\tilde\phi(t)}A_o\otimes\hat O\kket{\tilde\phi(t)}\nonumber\\
    &=\bbra{\tilde\phi(t)}(I+X)^{\otimes n_a}\otimes\hat O\kket{\tilde\phi(t)},
\end{align}
where $\kket{\tilde\phi(t)}=\big[\{\otimes_j R_j^z(\Omega t)\}\otimes \hat{I}\big]\kket{\phi}$. Thus, to measure the physical operator $\hat{O}$, we need to first apply time-dependent $Z$-rotations to the auxiliary qubits. We then measure $I+X$ on each of the auxiliary qubits, and we measure $\hat{O}$ on the physical qubits.
When we consider $t=nT$, then the stroboscopic expectation value is 
\begin{equation}\braket{\hat O(nT)}_{\phi}=\bbra{\phi}(I+X)^{\otimes{n_a}}\otimes \hat O \kket{\phi},
\end{equation}
and so the time-dependent $Z$-rotations are not needed in this case; instead we can measure $(I+X)^{\otimes{n_a}}\otimes \hat O$ directly on the extended Floquet state.
At all other times, we must first apply $Z$-rotations to the auxiliary qubits before making the measurements.
We see that measuring the expectation value of an operator is simplified by the fact that all the terms in $A_o=(I+X)^{\otimes n_a}$ bit-wise commute. Thus we can measure each auxiliary qubit in the $X$-basis, and there is no need for basis transformations involving entangling gates in the auxiliary space. We can measure $\hat{O}$ on the physical qubits using the standard procedure of first decomposing this operator in a Pauli basis and then measuring each term. Strategies for grouping Pauli strings into commuting sets can be used to reduce the total measurement count~\cite{gokhale2019minimizingstatepreparationsvariational,Verteletskyi2020,Crawford2021efficientquantum,Caratan_measurement_2021}. We show the quantum circuit to calculate the expectation value of an operator in Fig.~\ref{Fig:QC_Expt_Vals}.

\begin{figure}[h]
    \centering
    \resizebox{0.7\textwidth}{!}{ 
\begin{quantikz}
\lstick{$\ket{\bar{0}}$} & \gate[2]{\hat{\hat{U}}(\vec{\theta}_0 )}\qw & \gate{\bar{R}^z(\Omega t)}  &  \gate{\bar{H}}\qw & \qw  &\meter{} \\
\lstick{$\ket{\phi_\mathrm{p}}$} & \qw   &\qw & \gate[1]{\text{Basis change for $\hat O$}} &\qw & \meter{}\\
\end{quantikz}
    }
    \caption{Quantum circuit to calculate the time-dependent expectation value of an operator $\hat{O}$ using the extended Floquet state. Here $\ket{\bar{0}}$ represents the zeroth auxiliary state, while $\bar{H}$ and $\bar{R}^z(\Omega t)$ denote products of Hadamard gates and single-qubit $Z$-rotation gates, respectively, acting on the $n_a$ auxiliary qubits.}
    \label{Fig:QC_Expt_Vals}
\end{figure}

We emphasize that this method of calculating time-dependent expectation values is applicable to both near-term and fault-tolerant quantum computers, provided there is a method to prepare the extended Floquet state on the quantum computer. Additionally, this method of calculating the time-dependent expectation value in a Floquet state requires only a fixed-depth circuit, irrespective of the time within a period. In contrast, if we were to prepare a Floquet state \(\ket{\phi(0)}\) in the physical (or original) Hilbert space and use it to compute the time-dependent expectation value $\sandwich{\phi(0)}{\hat{U}^{\dagger}(t)\hat{O}\hat{U}(t)}{\phi(0)}$, we have to rely on Trotterization for \(\hat{U}(t)\). In this case, the circuit depth increases as the time within the period increases.








\begin{table}[t]
\centering
\small
\renewcommand{\arraystretch}{1.3}
\setlength{\tabcolsep}{3pt}

\begin{tabularx}{\linewidth}{|>{\raggedright\arraybackslash}X|
>{\centering\arraybackslash}X|
>{\centering\arraybackslash}X|
>{\centering\arraybackslash}X|}
\hline
\textbf{Feature}
& \textbf{Our work}
& \textbf{Ref.~\cite{Fauseweh2023} frequency-domain}
& \textbf{Ref.~\cite{Fauseweh2023} time-domain} \\
\hline

Need for Trotterization and IQPE & No & No & Yes \\
Valid for arbitrary truncation of aux. space & Yes & No & Not needed \\
Nature of algorithm & ADAPT-VQE & Hamiltonian-VQE & Time-domain VQE \\
Identification of a good initial state & Yes & No & No \\
Number of required ancillas & $O(\log_2 L)$ & $O(\log_2 L)$ & one for IQPE \\
Fixed-depth micromotion expectation values & Yes & No & No \\

\hline
\end{tabularx}

\caption{Comparison of our work with previous Floquet VQE algorithms~\cite{Fauseweh2023}.}
\label{tab:method_comparison}
\end{table}

\subsubsection{Fixed-depth preparation of time-periodic Floquet state in physical space}
Here, we show how to prepare the time-periodic Floquet state $\ket{\phi (t)}$ in the original, physical Hilbert space once we have prepared the extended Floquet state $\kket{\phi}$.  
Let $\kket{\phi}=\sum_{l=-N_c}^{N_c+1}\ket{l}\ket{\phi^{(l)}}$, where $\ket{\phi^{(N_c+1)}} =0$ as discussed in Eq.~\eqref{Eq:VQE_Ket_state} of the Appendix~\ref{App:Floq_Expt}. We know that 
\begin{align}
    \ket{\phi(t)} = \sum_{l=-N_c}^{N_c}e^{il\Omega t}\ket{\phi^{(l)}} = \left(\sum_{l=-N_c}^{N_c+1}\bra{l}\right)\left(e^{iA^d \Omega t} \otimes \hat{I}\right) \kket{\phi}.
\end{align}
We can represent $\frac{1}{\sqrt{2^{n_a}}}\sum_{l=-N_c}^{N_c+1}\bra{l}=\bra{+}^{\otimes n_a}$, and $\bigl(e^{iA^d \Omega t}\otimes \hat{I}\bigr)$ is a product of single-qubit $Z$-rotations up to a phase (as discussed in the previous section). This means that by applying these single-qubit $Z$-rotations and then projecting (i.e., measuring) the auxiliary qubits in the $X$ basis and postselecting the $\ket{+}^{\otimes n_a}$ outcome, we can prepare the time-periodic Floquet state at any time $t$ with success probability $1/2^{n_a}$. Since the auxiliary-space dimension grows only polynomially with the system size ($2^{n_a}=O(L^K)$), Floquet states at arbitrary times in the original Hilbert space can be prepared with a measurement cost that scales polynomially in the number of qubits. We provide the quantum circuit as shown in Fig.~\ref{Fig:Quantum_Circuit_State}.
\begin{figure}[H]
\centering
\resizebox{0.6\textwidth}{!}{
\begin{quantikz}
\lstick{$\vcenter{\hbox{$\ket{\bar{0}}$}}$}
  & \gate[2]{\hat{\hat{U}}(\vec{\theta}_0)}  & \gate{\bar{R}^z(\Omega t)} & \gate{\bar{H}} & \qw & \meter{} \\
\lstick{$\vcenter{\hbox{$\ket{\phi_\mathrm{p}}$}}$}
  & \qw &\qw &  \qw  & \qw & \qw \rstick{$\ket{\phi(t)}$} \\
\end{quantikz}
}
\caption{Quantum circuit that prepares a time-periodic Floquet state at any time $t$ using the extended Floquet state. Here, $\ket{\bar{0}}$ represents the zeroth auxiliary state, while $\bar{H}$ and $\bar{R}^z(\Omega t)$ denote products of Hadamard gates and single-qubit $Z$-rotations, respectively, acting on the $n_a$ auxiliary qubits. The final state $\ket{\phi(t)}$ is obtained probabilistically and therefore requires $O(\mathrm{poly}(L))$ measurements on the auxiliary register.}
\label{Fig:Quantum_Circuit_State}
\end{figure}

We note that the prepared Floquet state after measurement will differ from the exact state by a phase factor ($e^{i\Omega t/2}$). This is because $e^{iA^{\mathrm{d}}\Omega t}= e^{i\Omega t/2}\otimes_{j} R_j^z(\Omega t)= e^{i\Omega t/2} \bar{R}^z(\Omega t)$. 
Here $R_j^z(\Omega t)=e^{-i(2^{j-1}\Omega t)Z_j}$ is a single-qubit $Z$-rotation operator and $j\in\{0,1,\dots,n_a-1\}$. Thus, this phase can always be included in post-processing if needed.

We can contrast this with using $\hat{U}(T)$ to find the state as discussed in Ref.~\cite{Fauseweh2023}, where one can only obtain the Floquet state at a single reference time and quasienergy. Recovering the full time evolution of the Floquet state then requires additional time evolution of the state, and therefore the circuit depth increases with evolution time.

To compare our work with previous VQE-based approaches for periodically driven systems, we summarize the key features in Table~\ref{tab:method_comparison}. The comparison focuses on the representation of the Floquet problem, the nature of the variational algorithm, auxiliary-qubit requirements, initial-state preparation, and access to fixed-depth time-dependent observables. This highlights the broader scope of our framework.

\begin{figure}[t]
    \centering
\includegraphics[width=0.9\linewidth]{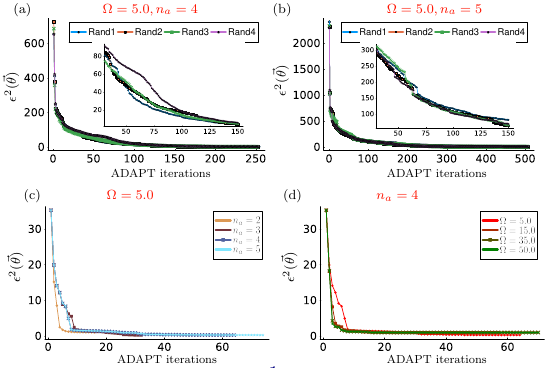}
    \caption{Squared Floquet energy, $\epsilon^2(\vec{\theta})$, of a periodically driven XYZ spin chain with magnetic field as a function of ADAPT iterations for various parameters. All parameters are expressed in units of the characteristic coupling constant $J_0$. The squared Floquet energy is shown for four random initial states (as written in the legends) for $\Omega = 5.0J_0$ and (a) $n_a = 4$ and (b) $n_a = 5$, respectively. The insets provide a closer look at the behavior from iterations 31 to 150. (c) The squared Floquet energy for four different values of $n_a$ at $\Omega = 5.0 J_0$. (d) The squared Floquet energy for four different values of $\Omega$ at $n_a = 4$. In both (c) and (d), the initial state is $\kket{\psi} = \ket{0}\otimes\ket{\phi_\mathrm{p}}$, where $\ket{\phi_\mathrm{p}} = \ket{\downarrow}^{\otimes L}$. The rest of the parameters are $L=3, $ $(\bar J_x,\bar J_y,\bar J_z)=(3.7, 2.8, 3.9)J_0, (\delta J_x,\delta J_y,\delta J_z)=(0.0,0.0,0.0)J_0$, $\bar B_z= 2.9 J_0,\delta B_z=2.7 J_0$. The couplings and magnetic field values are chosen arbitrarily aside from the conditions that they are of the same order of magnitude and do not exhibit any special relations. Other generic choices of parameters yield similar results. We use a mixed Floquet operator pool consisting of all possible Pauli operators on the auxiliary qubits and all weight-2 Pauli strings on the physical qubits.}
    \label{Fig:Initial_State}
\end{figure}
\section {Numerical simulation of driven XYZ model using Floquet ADAPT-VQE}\label{sec:Application}

\subsection{Model and simulations}
We now apply Floquet-ADAPT-VQE to study non-equilibrium physics in an XYZ spin chain with periodically driven couplings and with a time-periodic magnetic field. Both the couplings and the magnetic field are taken to have the same period $T=2\pi/\Omega$. The Hamiltonian is 
\begin{equation}
H(t) = \sum_{j=1}^{L-1} \Big( J_x(t)X_{j}X_{j+1} + J_y(t) Y_{j}Y_{j+1}+J_z(t) Z_{j}Z_{j+1}\Big) + \sum_{j=1}^{L} B_{z}(t) Z_j, 
\end{equation}
where 
\begin{align}
J_{\mu}(t) &= \bar{J}_\mu + \delta J_{\mu} \cos(\Omega t), \nonumber\\
B_z (t) &= \bar{B}_z + \delta B_z \cos(\Omega t),
\end{align}
and $\mu\in\{x,y,z\}$. 
The Fourier components of the Hamiltonian are
\begin{align}
    H^{(0)} &= \sum_{j=1}^{L-1} \Big( \bar{J}_x X_{j}X_{j+1} + \bar{J}_y Y_{j}Y_{j+1}+\bar{J}_z Z_{j}Z_{j+1}\Big) + \sum_{j=1}^{L}\bar{B}_{z} Z_j, \\
    H^{(1)} &= \sum_{j=1}^{L-1} \frac{1}{2}\Big(\delta J_x X_{j}X_{j+1} + \delta J_y Y_{j}Y_{j+1}+ \delta J_z Z_{j}Z_{j+1}\Big) + \sum_{j=1}^{L} \frac{1}{2}\delta B_{z} Z_j. 
\end{align}
Here, $H^{(1)}=H^{(-1)\dagger}=H^{(-1)}$. The extended Floquet Hamiltonian is given by
\begin{align}
    \hhat{H}_\mathrm{F}=  I \otimes \hat H^{(0)} + \Omega\big(A^{\mathrm{d}}\otimes  \hat I\big) +  (A^{(1)}+A^{(-1)})\otimes \hat H^{(1)} -(A_\mathrm{asy}^{(1)}+A_\mathrm{asy}^{(-1)})\otimes \hat H^{(1)}.
\end{align}

\begin{figure}[t]
    \centering
    \includegraphics[width=0.9\linewidth]
    {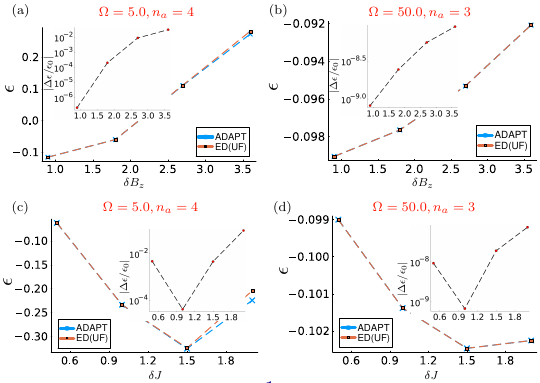}
    \caption{The quasienergies $\epsilon=\bbra{\psi_0}\hhat{H}_\mathrm{F} \kket{\psi_0}$ obtained using Floquet-ADAPT-VQE (``ADAPT'') as well as the exact quasienergies (``ED(UF)'') as a function of drive parameters in a periodically driven XYZ spin chain with a magnetic field. All parameters are expressed in units of the characteristic coupling constant $J_0$. The insets provide the relative error ($|\Delta\epsilon/\epsilon_0|=|(\epsilon-\epsilon_0)/\epsilon_0|$) of the Floquet-ADAPT-VQE quasienergies with respect to the exact quasienergies. (a,b) The quasienergies versus the driving amplitude of the magnetic field $\delta B_z$ when the couplings $J$ are static, i.e., $(\delta J_x, \delta J_y, \delta J_z)=(0.0, 0.0, 0.0) J_0$. In (a), we take 
    $\Omega = 5.0 J_0$ and $n_a = 4$ and in (b) $\Omega = 50.0 J_0$, $n_a =3$. (c,d) The quasienergies with drive terms applied to both the couplings and magnetic field, where the magnetic field amplitude is fixed at $\delta B_z = 2.7 J_0$ and the driven coupling terms are $(\delta J_x,\delta J_y,\delta J_z)=\delta J\times(1.9, 1.1, 1.2) J_0$. In (c), $\Omega = 5.0 J_0$ and $n_a = 4$ and in (d) $\Omega = 50.0 J_0$, $n_a =3$. The rest of the parameters for all panels are $L=4, $ $(\bar{J}_x,\bar{J}_y,\bar{J}_z)=(3.7, 2.8, 3.9) J_0$, $\bar{B}_z=2.9 J_0$. The initial state is $\kket{\psi_\mathrm{in}} = \ket{0}\otimes\ket{\phi_\mathrm{p}}$, where $\ket{\phi_\mathrm{p}} = \ket{+}^{\otimes L}$. We use a mixed Floquet operator pool consisting of all possible Pauli operators on the auxiliary qubits and all weight-2 Pauli strings on the physical qubits.}
    \label{Fig:QE_MeshdBz_MeshdJ}
\end{figure}
\begin{figure}[t]
    \centering
    \includegraphics[width=0.7\linewidth]{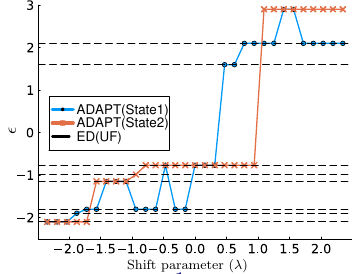}
    \caption{Quasienergy spectrum of a periodically driven XYZ spin chain as a function of the shift parameter $\lambda$ in the cost function. All parameters are expressed in units of the characteristic coupling constant $J_0$. The dashed lines show the exact quasienergies labeled by ED(UF). The $\text{State1}=\kket{\psi_1}$ and $\text{State2}=\kket{\psi_2}$ represent two initial states for the Floquet-ADAPT-VQE algorithm defined as $\kket{\psi_1} = \ket{0}\otimes\ket{\downarrow}^{\otimes L}$  and $\kket{\psi_2} = \ket{0}\otimes\ket{+}^{\otimes L}$.  The rest of the parameters are $L=3, n_a=4$, $(\bar{J}_x,\bar{J}_y,\bar{J}_z)=(3.7, 2.8, 3.9) J_0$, $(\delta J_x,\delta J_y,\delta J_z)=(1.9,1.1,1.2) J_0$, $\bar{B}_z= 2.9 J_0$, $\delta B_z=2.7 J_0$, and  $\Omega = 5.0 J_0$.  We use a mixed Floquet operator pool consisting of all possible Pauli operators on the auxiliary qubits and all weight-2 Pauli strings on the physical qubits.}
    \label{Fig:Full_Quasienrgy_Spectra}
\end{figure}



Before presenting the results of our simulations, we define the notation used in this section. The ADAPT quasienergy is defined as $\epsilon = \bbra{\psi(\vec{\theta}_0)}\hhat{H}_\mathrm{F}\kket{\psi(\vec{\theta}_0)}$, where $\kket{\psi(\vec{\theta}_0)}$ is an eigenstate of $\hhat{H}_\mathrm{F}^2$ obtained using Floquet-ADAPT-VQE. Additionally, we use the term ``squared Floquet energy'' to refer to $\epsilon^2(\vec\theta) = \bbra{\psi(\vec{\theta})}\hhat{H}_\mathrm{F}^2\kket{\psi(\vec{\theta})}$, where $\kket{\psi(\vec{\theta})}$ is a state in the  Floquet-Hilbert space at each ADAPT iteration. We compare the results of the Floquet-ADAPT-VQE algorithm with those obtained by the exact diagonalization (ED) of $\hat U_\mathrm{F}$, not with the ED of the extended Floquet Hamiltonian  $\hhat{H}_\mathrm{F}$. We do not use the ED of $\hhat{H}_\mathrm{F}$ because it is already approximate due to truncation. In our ED calculation, the eigenvalues $\epsilon_{\alpha}$ are obtained from $ \hat H_\mathrm{F} = (-i/T) \ln(\hat U_\mathrm{F})$, and we refer to them as the exact quasienergies. The time evolution operator $ \hat U(t)$ and the Floquet unitary $\hat U_\mathrm{F} = \hat U(T)$  are computed using Trotterization. We use the  L-BFGS optimizer to obtain our Floquet-ADAPT-VQE results.  Finally, in all that follows, the static and driving coupling parameters, as well as the magnetic fields, are expressed in the same units as the driving frequency. We will express all parameters in units of the characteristic coupling constant $J_0$.

Figure \ref{Fig:Initial_State} highlights the utility of selecting the auxiliary state as the eigenstate corresponding to the zero eigenvalue of the diagonal auxiliary matrix  $A^{\mathrm{d}}$. The figure shows the squared Floquet energy as a function of the number of ADAPT iterations. In Figs.~\ref{Fig:Initial_State}(a) and (b), we consider four random states for $n_a = 4$ and $n_a = 5$, respectively, at a driving frequency of $\Omega = 5.0 J_0$. The insets zoom in to show finer variations in squared energy over a portion of the ADAPT iterations. We observe that increasing $n_a$ results in a higher initial squared Floquet energy, which affects the convergence rate. Similarly, in Appendix~\ref{Appen:initial_state} we show that increasing $\Omega$ also leads to a higher initial squared Floquet energy. In Figs.~\ref{Fig:Initial_State}(c) and (d), we choose the initial auxiliary state to be the eigenstate of $A^{\mathrm{d}}$ with zero eigenvalue. In Fig.~\ref{Fig:Initial_State}(c), for a fixed driving frequency, we vary $n_a$ and observe that the initial squared Floquet energy does not depend on $n_a$. Similarly, varying the driving frequency in Fig.~\ref{Fig:Initial_State}(d), we see that the initial squared Floquet energy does not depend on the driving frequency either. These plots numerically confirm our proof, as discussed in Sec.~\ref{sec:Prep_Floq_state} and Appendix~\ref{Appen:initial_state}, that choosing the initial auxiliary state as the eigenstate of $A^{\mathrm{d}}$ with zero eigenvalue ensures that the initial squared energies are independent of $n_a$ and $\Omega$, leading to better convergence compared to using a random state. This choice of initial state becomes particularly important when studying larger systems at moderate or low driving frequencies, as these regimes require a larger number of auxiliary qubits.

Figure~\ref{Fig:QE_MeshdBz_MeshdJ} shows the quasienergies for different driving parameters at both moderate $O(\Omega) \approx O(\delta B, \delta J)$ and high driving frequencies ($\Omega \gg \delta B, \delta J$). Each panel includes an inset that displays the relative error in the quasienergy compared to the ED solution. Figures~\ref{Fig:QE_MeshdBz_MeshdJ}(a) and (b) show the variation in quasienergies with respect to the driven magnetic field for static coupling terms. In Fig.~\ref{Fig:QE_MeshdBz_MeshdJ}(a), we consider a moderate frequency ($\Omega = 5.0 J_0$) with $n_a = 4$ and observe that the Floquet-ADAPT-VQE quasienergies closely match the exact quasienergies, particularly at lower driving amplitudes. As the driving amplitude increases, the relative error also increases, suggesting that a higher number of auxiliary qubits is needed for improved accuracy. In Fig.~\ref{Fig:QE_MeshdBz_MeshdJ}(b), we examine a high frequency ($\Omega = 50.0 J_0$) with $n_a = 3$ and find that the Floquet-ADAPT-VQE and exact quasienergies are very close. Although the relative error does increase with the increase in amplitude, it remains well within acceptable limits for most applications. 
In Figs.~\ref{Fig:QE_MeshdBz_MeshdJ}(c) and (d), we consider the effects of driving terms in both the coupling and magnetic field. Here, we fix the driving amplitude of the magnetic field and vary the amplitude of the coupling terms. In Fig.~\ref{Fig:QE_MeshdBz_MeshdJ}(c), we analyze the variation in the quasienergy at a moderate frequency ($\Omega = 5.0 J_0$) with $n_a = 4$, while in Fig.~\ref{Fig:QE_MeshdBz_MeshdJ}(d), we examine it in the high-frequency limit ($\Omega = 50.0 J_0$) with $n_a = 3$. The results in these cases are qualitatively similar to those observed when only the magnetic field is driven. These figures demonstrate that systems with higher driving frequencies require fewer auxiliary qubits, while systems with lower frequencies need more auxiliary qubits for improved accuracy. Additionally, it is worth noting that results like those shown in Figs.~\ref{Fig:QE_MeshdBz_MeshdJ}(a) and (c) cannot be simulated accurately using the extended space method from Ref.~\cite{Fauseweh2023} due to its limited truncation capacity.
 
In Fig.~\ref{Fig:Full_Quasienrgy_Spectra}, we present the quasienergy spectrum in the central Floquet zone for the driven XYZ model with magnetic field obtained using numerical simulations of the Floquet-ADAPT-VQE algorithm. The dashed horizontal lines in the figure represent the exact quasienergies within the central Floquet zone. We use $\lambda$ as a tunable parameter in the cost function, Eq.~(\ref{Eq:Cost_function}), and discretize its values as $\lambda \in \{(j -2^{L+1})\Omega/2^{L+2}\}_{j=1}^{2^{L+2}-1}$. We begin by selecting the initial state $\kket{\psi_1} = \ket{0}\otimes\ket{+}^{\otimes 3}$ and calculate the Floquet-ADAPT-VQE quasienergies for the discretized values of $\lambda$. The blue line in the figure, representing the quasienergies for this case, captures all but one quasienergy at $\epsilon = -0.98375 J_0$. Next, we choose a different initial state, $\kket{\psi_2} = \ket{0}\otimes\ket{\downarrow}^{\otimes 3}$, and again calculate the Floquet-ADAPT-VQE quasienergies. This time, only half of the full quasienergies are captured, but it successfully identifies the quasienergy that was missed with the previous initial state.  We also note that in larger systems, where ED is not feasible, we may not know the specific locations of the missing quasienergies, but we can always determine how many are missing within the central Floquet zone. By finely discretizing $\lambda$, varying the initial state, and choosing the pool of operators wisely, we aim to obtain the full quasienergy spectrum. The primary advantage of this method is that each quasienergy calculation is independent of the others, leading to a reduced circuit depth compared to most excited-state VQE methods. In Appendix~\ref{Appen: Excited_state_VQE}, we use the Variational Quantum Deflation algorithm \cite{Higgott2019variationalquantum} and plot the resultant quasienergies.

\begin{figure}[t]
    \centering
    \includegraphics[width=0.9\linewidth]{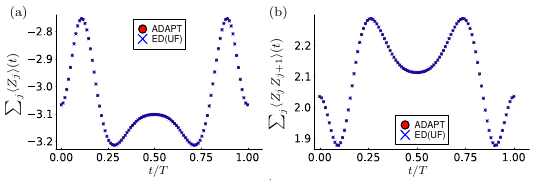}
    \caption{Expectation values of operators with respect to a Floquet state as a function of time, calculated using Floquet-ADAPT-VQE and direct Trotterization of $\hat U(t)$. All parameters are expressed in units of the characteristic coupling constant $J_0$. (a) The expectation value of the operator $\sum_j Z_j$. (b) The expectation value of $\sum_j Z_{j}Z_{j+1}$. Results for both expectation values are shown over one driving period. For the Floquet-ADAPT-VQE calculation, we numerically simulate the algorithm to compute the expectation value of an operator in the extended Floquet state. The extended Floquet state is obtained numerically using Floquet-ADAPT-VQE. The rest of the parameters are $L=4, (\bar J_x,\bar J_y,\bar J_z)=(3.7, 2.8, 3.9) J_0, (\delta J_x,\delta J_y,\delta J_z)=(1.9,1.1,1.2) J_0$, $\bar B_z= 2.9 J_0,\delta B_z=2.7 J_0$. The initial state is $\kket{\psi_\mathrm{in}} = \ket{0}\otimes\ket{\phi_\mathrm{p}}$, where $\ket{\phi_\mathrm{p}} = \ket{+}^{\otimes L}$. We use a mixed Floquet operator pool consisting of all possible Pauli operators on the auxiliary qubits and all weight-2 Pauli strings on the physical qubits. }
    \label{Fig:Op_Expt}
\end{figure}

Figure~\ref{Fig:Op_Expt} shows the expectation values of two operators. In Fig.~\ref{Fig:Op_Expt}(a), we show the expectation value of the magnetization ($\sum_j Z_j$), while Fig.~\ref{Fig:Op_Expt}(b) shows the expectation value of the two-point spin-$Z$ operator ($\sum_j Z_j Z_{j+1}$). These calculations were performed using the algorithms developed in this paper. We first employed Floquet-ADAPT-VQE to prepare the extended Floquet state, then used our algorithm to calculate the expectation values in the Floquet-Hilbert space. For comparison, we also computed the exact time evolution of the operators by applying Trotterization to the time evolution operator $\hat U(t)$ over one period. Both methods produce identical results, clearly demonstrating the periodic nature of the expectation values.

\begin{figure}
\includegraphics[width=0.45\textwidth]{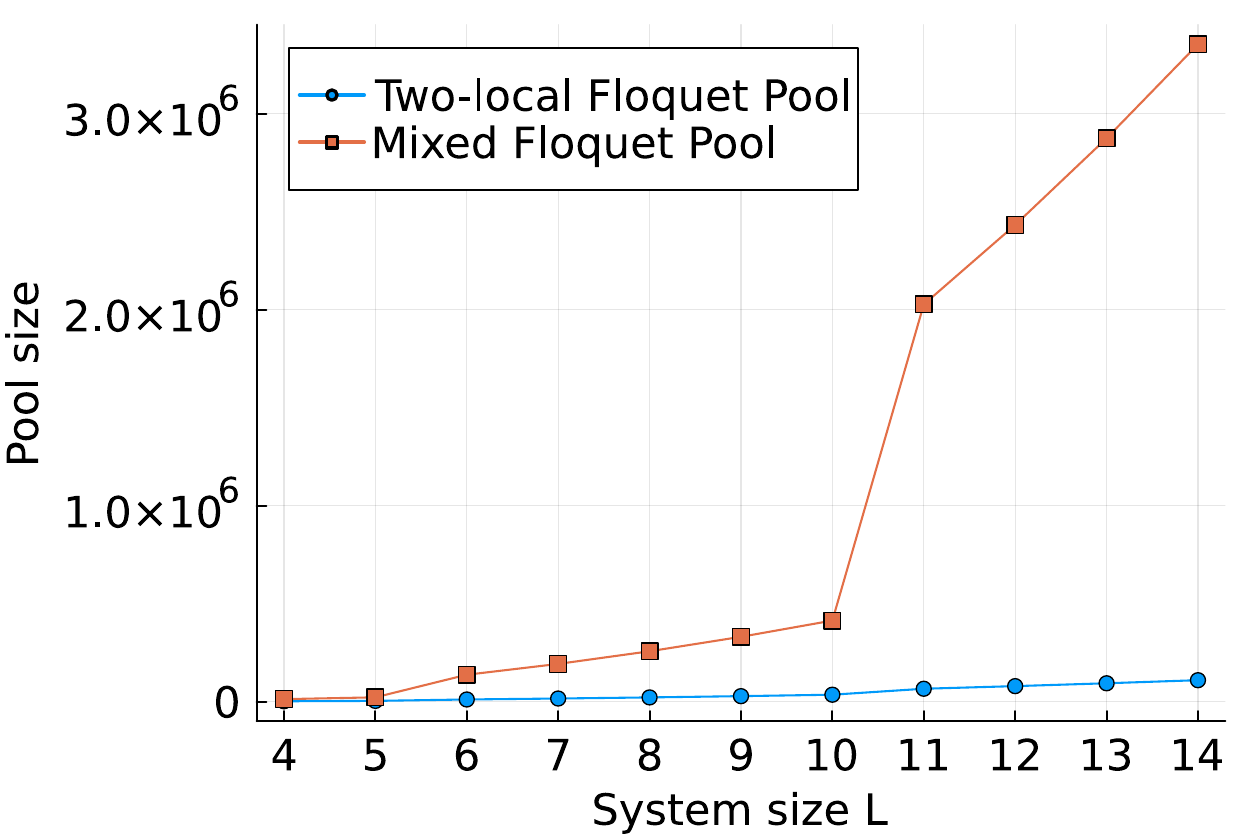}
\includegraphics[width=0.45\textwidth]{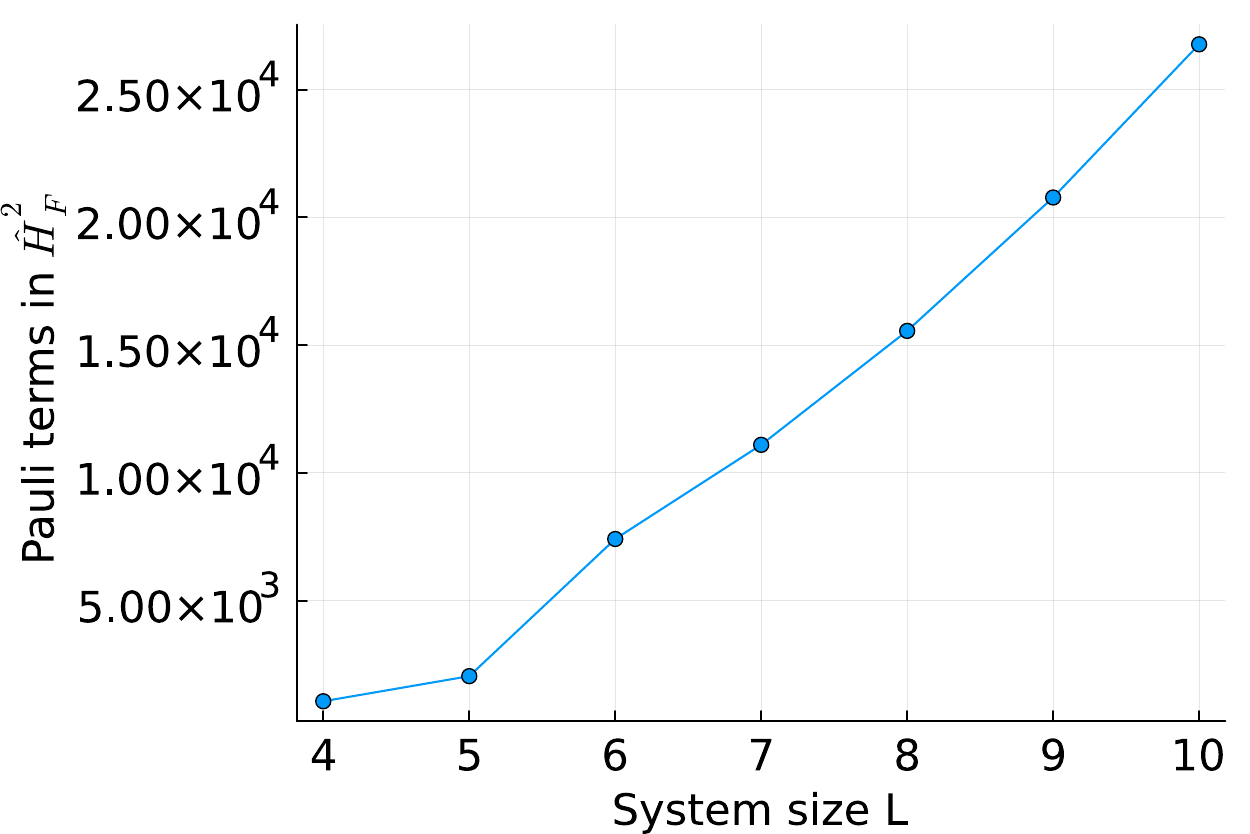}
  \caption{The number of Pauli strings in the pool and in the squared Floquet Hamiltonian as a function of $L$. The left panel shows the pool size with respect to system size. The right panel shows Pauli terms (number of Pauli strings) in the squared extended Floquet Hamiltonian. The two-local Floquet pool is composed of all tensor products between weight-2 Pauli strings on the auxiliary subspace and weight-2 Pauli strings on the physical subspace. In contrast, the mixed Floquet pool includes arbitrary Pauli strings on the auxiliary subspace and only weight-2 Pauli strings on the physical subspace.}
  \label{fig:Pauli}
\end{figure}

    \subsection{Resource estimation for the driven XYZ model}
    Here, we estimate the number of auxiliary qubits required for the auxiliary space, as well as the scaling of the number of Pauli strings appearing in the operator pool and in the squared extended Floquet Hamiltonian, for the driven XYZ model. Based on the truncation analysis in Appendix~\ref{Appen:truncation}, the number of auxiliary qubits scales as
\begin{align}
    n_a &= \left\lceil \log_2 (L)  + \log_2\left(\frac{8J_{\max}}{\Omega}\right) \right\rceil ,
\end{align}
where $J_\mathrm{max}$ is the largest interaction strength in the Hamiltonian.

 We first consider a mixed Floquet operator pool ($\mathcal{P}$) consisting of all possible Pauli operators on the auxiliary qubits and all weight-2 Pauli strings on the physical qubits. The total number of operators ($\mathcal{N}(\mathcal{P})$) then scales as
\begin{align}
    \mathcal{N}(\mathcal{P})= 4^{n_a}\times 9L(L-1)/2\sim O\left(\left(\frac{J_{\textrm{max}}}{\Omega}\right)^2 L^4\right).
\end{align}
Alternatively, if we use a restricted mixed Floquet operator pool ($\mathcal{P}$) consisting  of all weight-2 Pauli string operators on both the auxiliary and physical qubits, then 
\begin{align}
    \mathcal{N}(\mathcal{P})= [9n_a(n_a-1)/2]\times [9L(L-1)/2]\sim O\left( \left(\log_2\left(\frac{J_{\textrm{max}}}{\Omega}\right)+ \log_2L\right)^2 L^2\right).
\end{align}

For the squared extended Floquet Hamiltonian, the number of Pauli strings scales as  
\begin{align}
    \mathcal{N}(\hhat{H}_F^2)\sim O\Big(4^{n_a}\times L^2\Big)\sim O\Big(L^4\Big).
\end{align}
Thus, the energy-measurement cost scales as $O(L^4)$, while the gradient-measurement cost scales as $O(\log_2^2(L)L^6)$ for the restricted mixed Floquet pool consisting of all weight-2 Pauli strings acting separately on the auxiliary and physical spaces, assuming $J_{\max}/\Omega$ is fixed.

   In this section, we focus on the scaling of the number of Paulis in the cost function and operator pool. These affect the complexity of our numerical simulations. However, in hardware implementations, grouping techniques can be employed to reduce the measurement overhead, as in Appendix~\ref{Appen:Meaurement_Scaling_of_cost_fuction}.

    In Fig.~\ref{fig:Pauli}, we show the number of Pauli strings as a function of the number of qubits, or equivalently the system size. The left panel shows the operator-pool size with system size. As expected from the theoretical analysis, the mixed Floquet pool consisting of all weight-2 Pauli strings on both the auxiliary and physical parts grows much more slowly than the mixed Floquet pool consisting of all weight-2 Pauli strings on the physical part and all Pauli strings on the auxiliary part. The right panel shows the number of Pauli terms in the squared extended Floquet Hamiltonian.

\section{Conclusions and Outlook}\label{sec:Discussion}

In this work, we presented a general framework for studying periodically driven many-body quantum systems on a digital quantum computer. We leveraged the Floquet-Hilbert space---a composite vector space comprised of auxiliary and physical spaces---to make the Hamiltonian time-independent. We showed how to map this Hamiltonian in terms of Pauli strings to implement it on a quantum processor. We introduced a hybrid algorithm that we call Floquet-ADAPT-VQE. We defined a cost function based on the square of the shifted, extended Floquet Hamiltonian, enabling us to prepare the extended Floquet state and identify quasienergies anywhere within a Floquet zone. A randomly chosen initial state will typically have low overlap with the target state, leading to less efficient convergence. To address this, we demonstrated how choosing an initial state on the auxiliary space as an eigenstate of a diagonal auxiliary matrix with zero eigenvalue is a good starting point, independent of the number of auxiliary qubits and the driving frequency. Additionally, we provided a protocol for calculating the expectation value of an operator in a Floquet state.

To showcase the utility of Floquet-ADAPT-VQE, we performed classical numerical simulations of our algorithm applied to a periodically driven XYZ spin chain with a magnetic field. Our simulations numerically verified our assertion that the squared Floquet energy of a random initial state remains significantly above the desired Floquet zone ($0 < \epsilon^2 <\Omega^2 / 4$) as the number of auxiliary qubits and driving frequency increase, leading to a slower convergence of the algorithm. In contrast, the squared Floquet energy of our carefully chosen initial state remains independent of the auxiliary qubits and driving frequency, resulting in better convergence and fewer Floquet-ADAPT-VQE iterations. We also examined how the quasienergies depend on the driving amplitude at both high and moderate frequencies. Furthermore, we calculated the full quasienergy spectrum of the central Floquet zone by scanning over the shift parameter $\lambda$ in the cost function. Finally, we calculated the expectation value of the net magnetization $(\sum_j Z_j)$ and the sum of the two-point $Z$-operator $(\sum_j Z_j Z_{j+1})$ in a Floquet state using our algorithms, confirming the efficacy of our protocol for computing expectation values in a Floquet state.

We would like  to emphasize that although we used a driven XYZ Hamiltonian as an example, our algorithm is applicable to any type of time-periodic Hamiltonian, including fermionic Hamiltonians. The algorithm is based on transforming a periodic Hamiltonian into an extended Floquet Hamiltonian, which provides a time-independent description that is approximate when a finite number of auxiliary qubits is used. With an arbitrary number of auxiliary qubits, we can approximate any type of periodically driven system. However, with a limited number of auxiliary qubits, the approximation is most effective for systems with a continuous periodic drive featuring a few frequency modes and operating at moderate to high frequencies. Consequently, we believe that our algorithm will be particularly useful for studying phenomena related to light-matter interactions in many-body quantum systems.

Although our study primarily focuses on Floquet-ADAPT-VQE, some of the results obtained in this paper have broader applications across various types of quantum simulation methods. For example, choosing an initial auxiliary state as an eigenstate of the diagonal auxiliary matrix with a zero eigenvalue can reduce the number of convergence steps in any variational algorithm where the cost function is defined using extended Floquet Hamiltonian. Moreover, the Pauli decomposition of auxiliary matrices can be beneficial for any quantum simulation that uses the extended Floquet Hamiltonian. This could also be used for simulating  quantum dynamics for a time-periodic Hamiltonian \cite{Levante_Floq_Dynamics} like  studying  dynamics of light-driven matter \cite{Vega_PRR_2022}. Additionally, our protocol for calculating the expectation value of an operator can be applied to any gate-based quantum computer once the extended Floquet state is prepared.

Before concluding, we briefly outline two phenomena that could be explored in future studies using our formalism. First, our approach can be used to detect time crystal behavior in a system. The eigenstates of the Floquet unitary in time crystals are long-range correlated states that come in pairs (in the case of period-doubled time crystals), with a quasienergy difference of $\Omega/2$. After finding the first eigenstate ($\epsilon_0< 0$) using Floquet-ADAPT-VQE, the corresponding pair can be identified by tuning $\lambda = \epsilon_0+\Omega/2$. Additionally, our algorithm can also be employed to calculate the stroboscopic time evolution of an operator in the case of a time crystal. More details can be found in Appendix~\ref{App:TC_detection}.

Secondly, we can study phenomena related to light-matter interactions in many-body systems. This algorithm is suitable for simulating multi-frequency light-matter interactions in such systems. By leveraging the extended Floquet state, we can investigate the linear response in periodically driven many-body interacting systems, similar to the theoretical studies conducted in non-interacting tight-binding Hamiltonians \cite{Kumar_PRB_2020}. Furthermore, this approach can be applied to study high harmonic generation in many-body interacting solids \cite{Kumar_HHG_2022arXiv}.


     Our work establishes a foundation for simulating time-periodic Hamiltonians within the Floquet--Hilbert-space formalism. With our efficient truncation scheme, Floquet-ADAPT-VQE requires an ancillary (auxiliary) qubit overhead that scales only logarithmically with system size, while the number of measurements scales polynomially, as is typically the case for variational algorithms. In particular, we show that for a time-periodic Hamiltonian with a single Fourier mode and geometrically local Pauli strings of maximum weight $w$, the measurement cost for the cost function scales as $O(\mathrm{polylog}(L)L)$, while the gradient-measurement cost scales as $O(\mathrm{polylog}(L)L^3)$ when the operator pool consists of all weight-2 Pauli strings chosen separately on the physical and auxiliary spaces. This favorable scaling follows from two ingredients: the geometrically local Pauli-string structure and the efficient grouping, proven in Appendix B, of the Pauli strings appearing in the auxiliary matrices for Fourier modes $r=2^q$, where $q$ is a non-negative integer.

As with all VQEs, the measurement overhead renders near-term hardware demonstrations for larger system sizes challenging, as does the need for efficient measurement-feedback loops required by adaptive algorithms such as ours. Encouragingly, significant progress has been made in reducing the measurement cost  of static ADAPT-VQE~\cite{anastasiou2023reallymeasureoperatorgradients,Ramoa2024,ramôa_2024_CEOPool} relative to its original formulation. In fact, ADAPT often requires far fewer measurements than the best non-adaptive algorithms because it avoids the need for many random parameter initializations~\cite{ramôa_2024_CEOPool}. Thus, the additional measurement cost associated with pool gradients may not be the dominant bottleneck compared to non-adaptive VQE. We expect similar advantages for certain classes of time-periodic Hamiltonians. While some of these ideas extend to Floquet-ADAPT-VQE, developing strategies tailored to the Floquet setting remains an important open direction. Promising avenues include improved cost functions that exploit the linear structure of the extended Floquet Hamiltonian $\hat{\hat{H}}_F$ and more effective operator pools for specific problem classes.

 In conclusion, our work establishes a foundation for developing future quantum simulation algorithms for time-periodic systems using the extended Floquet Hamiltonian. Specifically, our Floquet-ADAPT-VQE algorithm facilitates the determination of quasienergies and the preparation of Floquet states within the Floquet-Hilbert space, thereby advancing the study of non-equilibrium physics in periodically driven many-body systems on quantum processors. To implement our algorithm on NISQ devices, further efforts are required to identify more efficient (smaller) operator pools and to study the algorithm's performance under various noise models, which we leave for future work.

\section{Data availability}
The code and data used to generate the simulations and figures can be found at \url{https://github.com/AbhiPhy2020/Floq_ADAPT_v1.jl}, and make use of the ADAPT-VQE code available at \url{https://github.com/kmsherbertvt/publiccopy-bp-mitigation-code}. The code requires NLOpt version 0.65.
The Exact Diagonalization code can be found at \url{https://github.com/AbhiPhy2020/Floq_ED_v0.jl}.

\acknowledgements
A.K. and K.S. would like to thank Samantha Barron and Kyle Sherbert for their helpful comments and discussion on ADAPT-VQE code. K. S. and S. E. E. were supported by the U.S. Department of Energy, Office of Science, National Quantum Information Science Research Centers, Co-design Center for Quantum Advantage (C$^2$QA) under Contract No. DE-SC0012704. E.B. and N.M. acknowledge support by the U.S. Department of Energy, Office of Science, Office of Advanced Scientific Computing Research, under Award Number DE-SC0025430.

\section*{Author contributions}
A. Kumar conceived the project and led the development of the theoretical framework and algorithmic approach. K. Shirali and A. Kumar performed the numerical simulations and analysis. E. Barnes contributed to the theoretical development by refining the formulation and methodology. S. E. Economou, N. J. Mayhall, and E. Barnes contributed to the conceptual development through discussions and feedback. All authors contributed to the writing and revision of the manuscript.

\bibliographystyle{quantum}
\bibliography{Ref}

\newpage
\appendix
\section{Extended Floquet Hamiltonian}\label{Appen:Block_Extd_Floq_Ham}
 As described in the main text, the block-diagonalized extended Floquet Hamiltonian lives in a total space that includes the physical space and an $n_a$-qubit auxiliary space. We choose the Fourier indices to range from $-N_\mathrm{c}$ to $N_\mathrm{c}+1$. Here is an example of an extended Floquet Hamiltonian in the case of a three-qubit auxiliary space and for a driving term that contains 5 Fourier modes:
 \begin{equation}
 \begin{pmatrix}
   \hat{H}^{(0)}-3\Om  &  \hat{H}^{(-1)} &  \hat{H}^{(-2)}    & 0  & 0 & 0  &0  &0\\
     \hat{H}^{(1)}&  \hat{H}^{(0)}-2\Om  &  \hat{H}^{(-1)} &  \hat{H}^{(-2)}      & 0  & 0 &0 &0\\
     \hat{H}^{(2)} &   \hat{H}^{(1)} &  \hat{H}^{(0)}-\Om  &  \hat{H}^{(-1)} &  \hat{H}^{(-2)}  & 0 &0 &0 \\
      0 &  \hat{H}^{(2)}    &  \hat{H}^{(1)} &  \hat{H}^{(0)} &   \hat{H}^{(-1)}  &  \hat{H}^{(-2)}  &0 &0\\
     0 & 0    &  \hat{H}^{(2)}  &  \hat{H}^{(1)} &  \hat{H}^{(0)}+\Om  &  \hat{H}^{(-1)} & \hat{H}^{(-2)}  &0   \\
       0 &  0  &  0  &  \hat{H}^{(2)}   &   \hat{H}^{(1)} &  \hat{H}^{(0)}+2\Om  & \hat{H}^{(-1)}  & 0\\
       0&  0 & 0  &  0  &  \hat{H}^{(2)}   &   \hat{H}^{(1)} &  \hat{H}^{(0)}+3\Om & 0 \\
        0&0&0 & 0  &  0  &  0   &   0 &  \hat{H}^{(0)}+4\Om 
  \end{pmatrix}.\label{eq:5mode_example}
\end{equation}
We see that the largest-energy Floquet zone has been separated out and does not mix with other zones even after taking any power of $\hhat{H}_\mathrm{F}$.


Before moving forward, we would like to clarify the subtle distinction between the Fourier basis and the Floquet-zone basis. The Fourier basis enters because we expand the Floquet state in Fourier modes in the original Hilbert space. The Floquet-zone basis is obtained from this Fourier basis by block-diagonalizing the extended Floquet Hamiltonian.  There always exists a block-diagonalizing unitary matrix ($\hhat{U}_{\textrm{BD}}$) \cite{Eckardt_High_Freq} such that
\begin{align}\label{Eq_Appen:Block_diagonalization}
    \hhat{U}_{\textrm{BD}}\hhat{H}_F \hhat{U}_{\textrm{BD}}^{\dagger} = I\otimes\hat{H}_F + \Omega A^\textrm{d}\otimes \hat{I}.
\end{align}
Here, we note that $\hat{H}_F $ is the Floquet Hamiltonian in the physical space. Strictly speaking, the above relation is exact only in the infinite-dimensional auxiliary space. For a sufficiently large finite truncation, the corresponding diagonalization holds approximately, although exact discrete time translation symmetry (shift by $n\Omega$) is lost. Nevertheless, for the purpose of clarifying the relation between the two bases, we write the block-diagonalization of the  matrix in Eq.~\eqref{eq:5mode_example} as:
 \begin{align}
 \hhat{U}_{\textrm{BD}}\hhat{H}_F \hhat{U}_{\textrm{BD}}^{\dagger} &\approx \nonumber\\
 &\hspace{-0.5in}
 \begin{pmatrix}
  \hat{H}_F-3\Om  & 0 &  0      & 0  & 0 & 0  & 0 & 0 \\
   0 &\hat{H}_F-2\Om  & 0 &  0      & 0  & 0 & 0  & 0\\
    0 & 0 &  \hat{H}_F-\Om  &  0 & 0  & 0 & 0  & 0   \\
     0 &0    &  0 &  \hat{H}_F &   0 & 0  & 0  & 0 \\
      0 & 0 &  0 &  0 &  \hat{H}_F+\Om  &  0  & 0  & 0   \\
         0 &0 &    0  &  0  &   0&  \hat{H}_F+2\Om & 0  & 0    \\
        0  &  0  &   0  &  0  &  0  &   0&  \hat{H}_F+3\Om  &  0    \\
       0   &  0 &  0  &     0  &  0  &  0  &   0&  \hat{H}^{(0)}+4\Om    \\
  \end{pmatrix} .
\end{align}
We note that the term in the last block row and block column, $(\hat{H}^{(0)}+4\Omega)$, is irrelevant, while all other blocks represent different Floquet zones. By finding the eigenvalues of $\hat{H}_F$, we obtain the full quasienergy spectrum. If one follows this two-step diagonalization process, one obtains the quasienergies, but access to the Floquet states is restricted to stroboscopic times $t = nT$, with $n \in \mathbb{Z}^+ \cup \{0\}$. However, if we diagonalize the extended Floquet Hamiltonian directly, we obtain not only the quasienergies but also the Fourier modes of the Floquet states, which can be used to determine their properties at arbitrary times.  This is crucial for obtaining fixed-depth time-dependent expectation values of arbitrary observables in a Floquet state, as well as for preparing the Floquet state at arbitrary times using a fixed-depth circuit and measurements, as discussed further in Sec.~\ref{sec: Fixed_depth_obs_state}.

Now, we focus on the auxiliary part of the extended Floquet Hamiltonian. We express the auxiliary matrix and auxiliary vectors in terms of the standard matrix basis. A ket vector $\ket{n}$ in the $n_a$-qubit auxiliary space (of dimension $2^{n_a}= 2(N_\mathrm{c}+1)$) can be represented in terms of the standard basis vectors as:
\begin{equation}
 \ket{n}=e_{N_{\mathrm{c}}+n+1},
\end{equation}
where $-N_{\mathrm{c}}\leq n \leq N_{\mathrm{c}}+1$ and $e_{N_{\mathrm{c}}+n+1}\in\{e_1,e_2,\dots,e_{2^{n_a}}\}$ is standard basis (column) vector in $\mathbb{R}^{2^{n_a}}$ with a single non-zero entry of 1 at position $N_{\mathrm{c}}+n+1$.\ Similarly, the bra vector $\bra{n}$ corresponds to a standard basis row vector.\\
Likewise, auxiliary matrices \(A^{(|r|)}\), defined as 
\begin{equation}
A^{(|r|)} = \sum_{n=-N_{\mathrm{c}}}^{N_{\mathrm{c}}+1-|r|} \ket{n + |r|}\bra{n},
\end{equation}
can also be expressed in terms of the standard matrix basis in \(\mathbb{R}^{2^{n_a}} \times \mathbb{R}^{2^{n_a}}\). Specifically, we can write:
\begin{align}
\ket{n}\bra{m} &= E_{N_{\mathrm{c}} +n+1, N_{\mathrm{c}}+m+1},
\end{align}
where $n, m \in [-N_{\mathrm{c}}, N_{\mathrm{c}} + 1]$ and  $E_{N_{\mathrm{c}} + n+1, N_{\mathrm{c}} + m+1}\in\{E_{1,1},E_{1,2},\dots, E_{1,2^{n_a}},\dots,E_{2^{n_a},2^{n_a}}\}$ is a standard matrix basis element in \(\mathbb{R}^{2^{n_a}} \times \mathbb{R}^{2^{n_a}}\). This matrix has a single non-zero entry of 1 at the position $(N_{\mathrm{c}} + n+1, N_{\mathrm{c}} + m+1)$. We also have $A^{(-|r|)}= A^{(|r|)\dagger}$.
%
Now, we give explicit expressions of auxiliary ket vectors as column vectors for $n_a=3$ qubits:
\begin{align}
    \ket{-3}&=(1,0,0,0,0,0,0,0)^\mathrm{T}, \ket{-2}=(0,1,0,0,0,0,0,0)^\mathrm{T}, \ket{-1}=(0,0,1,0,0,0,0,0)^\mathrm{T},  \nonumber\\ 
    \ket{0}&=(0,0,0,1,0,0,0,0)^\mathrm{T},
    \ket{1}=(0,0,0,0,1,0,0,0)^\mathrm{T}, \ket{2}=(0,0,0,0,0,1,0,0)^\mathrm{T}, \nonumber\\  
    \ket{3}&=(0,0,0,0,0,0,1,0)^\mathrm{T}, \ket{4}=(0,0,0,0,0,0,0,1)^\mathrm{T},
\end{align}
and auxiliary bra vectors are row vectors (same as above without transpose). \\
Now the auxiliary matrix for $r=1$ is:  
\begin{align}
    A^{(1)}&= \sum_{n=-3}^{3} \ket{n +1}\bra{n} 
    = \begin{pmatrix}
    0  &  0 & 0 & 0  & 0 & 0  & 0  & 0\\
    1  &  0  & 0 &  0  & 0  & 0 & 0 & 0\\
    0 &  1 &  0  &  0 &  0  & 0 &0 &0 \\
      0 & 0    &  1 &  0 &  0  &  0  &0 &0\\
     0 & 0    &  0  &  1 &  0  &  0 & 0  &0   \\
       0 &  0  &  0  &  0  &   1 &  0  & 0  & 0\\
       0&  0 & 0  &  0  &  0  &   1 &  0 & 0 \\
        0&0&0 & 0  &  0  &  0   &   1 &  0
  \end{pmatrix},
\end{align}
and $A^{(-1)}=A^{(1)\dagger}$. Similarly, the asymmetric auxiliary matrix for $r=1$ is:
\begin{flalign}
    A_{\mathrm{asy}}^{(1)}&=\ket{4}\bra{3} 
    = \begin{pmatrix}
    0  &  0 & 0 & 0  & 0 & 0  & 0  & 0\\
    0  &  0  & 0 &  0  & 0  & 0 & 0 & 0\\
    0 &  0 &  0  &  0 &  0  & 0 &0 &0 \\
      0 & 0    &  0 &  0 &  0  &  0  &0 &0\\
     0 & 0    &  0  &  0 &  0  &  0 & 0  &0   \\
       0 &  0  &  0  &  0  &   0 &  0  & 0  & 0\\
       0&  0 & 0  &  0  &  0  &   0 &  0 & 0 \\
        0&0&0 & 0  &  0  &  0   &   1 &  0
  \end{pmatrix}.
\end{flalign}

\subsection{Scaling of Pauli strings for off-diagonal matrices in terms of auxiliary qubits}\label{Appen:Cal_Paui_Off}
In Sec.~\ref{subsec:Auxi_Algo_Pauli} of the main text, we  provide the decomposition of off-diagonal auxiliary matrices into Pauli strings.
Here we calculate the scaling of the number of Pauli strings in these decompositions with respect to the number of auxiliary qubits $n_a$. For a given $r$, we calculate $k_0=\lceil \log_2 (|r|)\rceil$, and we can write the auxiliary matrix for $n_a$ qubits as:
\begin{align}\label{Eq:Aux_Mat_r}
    A^{(-|r|)}_{n_a} = I^{\otimes(n_a-k_0)}\otimes A_{k_0}^{(-|r|)}+\sum_{k=0}^{n_a-k_0-1} I^{\otimes k}\otimes P^{+}\otimes(P^{-})^{\otimes(n_a-k-k_0-1)} \otimes Q^{(-|r|)}_{k_0}, \\
     A^{(|r|)}_{n_a} = I^{\otimes(n_a-k_0)}\otimes A_{k_0}^{(|r|)}+\sum_{k=0}^{n_a-k_0-1} I^{\otimes k}\otimes P^{-}\otimes(P^{+})^{\otimes(n_a-k-k_0-1)}\otimes Q^{(|r|)}_{k_0},
\end{align}
where $ A_{k_0}^{(r)}$ and $Q_{k_0}^{(r)}$  are two unknown quantities corresponding to $r=\pm |r|$. We can calculate the number of Pauli strings for $A^{(r)}_{k_0}$ and $Q^{(r)}_{k_0}$ using Eqs.~\eqref{eq:A1}--\eqref{eq:Aasy} of Sec.~\ref{subsec:Auxi_Algo_Pauli}. After that,  we can calculate the total number of Pauli strings for any auxiliary matrices under arbitrary qubit truncation. We denote the number of Pauli strings in $A_{k_0}^{(r)}$ and $Q_{k_0}^{(r)}$ as  $\mathcal{N}_\mathrm{P}\Bigl(A_{k_0}^{(r)}\Bigr)$ and $\mathcal{N}_{\mathrm{P}}\Bigl(Q_{k_0}^{(r)}\Bigr)$ respectively. The total number of Pauli strings is:
\begin{align}
    \mathcal{N}_\mathrm{P}\Bigl(A^{(r)}_{n_a}\Bigr)&= \mathcal{N}_\mathrm{P}\Bigl(A^{(r)}_{k_0}\Bigr)+ \mathcal{N}_\mathrm{P}\Bigl(Q^{(r)}_{k_0}\Bigr)\sum_{n=1}^{n_a-k_0} 2^{n}\nonumber\\
    &=\mathcal{N}_\mathrm{P}\Bigl(A^{(r)}_{k_0}\Bigr)+ \mathcal{N}_\mathrm{P}\Bigl(Q^{(r)}_{k_0}\Bigr)2^{n_a-k_0+1}\Bigl(1-(1/2)^{n_a-k_0}\Bigr).
\end{align}
Here, $\mathcal{N}_\mathrm{P}\Bigl(A_{k_0}^{(r)}\Bigr)$ and $\mathcal{N}_\mathrm{P}\Bigl(Q_{k_0}^{(r)}\Bigr)$ depend on $r$ or $k_0$ only and not on $n_a$.
 In particular, for $r=\pm 2^{k_0}$, we have $A^{(r)}_{k_0}=0$ and $Q^{(r)}_{k_0}=I^{\otimes k_0}$. So, $\mathcal{N}_\mathrm{P}(A^{(r=\pm 2^{k_0})}_{n_a})= 2^{n_a-k_0+1}\bigl(1-(1/2)^{n_a-k_0}\bigr)\leq 2^{n_a-k_0+1} $. In general, for $n_a\gg k_0 $, we have $ \mathcal{N}_\mathrm{P}\Bigl(A^{(r)}_{n_a}\Bigr)\approx O(2^{n_a})$.

\subsection{Efficient truncation and resulting size of the auxiliary space} \label{Appen:truncation}
In this section, we discuss the efficient truncation of the auxiliary space using the parameters of the time-periodic Hamiltonian. To do this, we follow the argument given in Sec. IIB of Ref.~\cite{rudner2020FloquetEngineersHandbook}. The extended Floquet Hamiltonian is closely analogous to a hopping Hamiltonian with an electric field on a one-dimensional lattice. As a result, due to a phenomenon analogous to Stark localization, the wave function (extended Floquet state) becomes localized in the Fourier basis. We rewrite the extended Floquet Hamiltonian as
\begin{align}
    \hhat{H}_F = \hhat{H}_{F0} +\Omega A^\textrm{d},
\end{align}
where $\hhat{H}_{F0}$ is an effective kinetic term in the Fourier basis, and $\Omega A^\textrm{d}$ is like an electric field potential.
The localization length in the Fourier basis is $\tilde{\ell}\approx \mathcal{W}/\Omega$, where $\mathcal{W}$ represents the many-body bandwidth of $\hat{\hat{H}}_{F0}$. Beyond this length, the Fourier modes of a Floquet state are exponentially suppressed.
Therefore, for the truncation, $n_a$ should satisfy 
\begin{align}\label{Eq:Truncation_na}
    2^{n_a} \gtrsim \mathcal{W}/\Omega.
\end{align}
Now, we estimate an upper bound on the many-body bandwidth ($\mathcal{W}$) of $\hat{\hat{H}}_{F0}$.
Using the spectral norm (the largest absolute eigenvalue of a Hermitian operator) of $\hat{\hat{H}}_{F0}$, we obtain the following upper bound on the many-body bandwidth:
\begin{align}
    \mathcal{W} \leq 2\|\hhat{H}_{F0}\|_2\leq 2\Big(\|I\otimes \hat{H}^{(0)}\|_2 + \sum_{r\in \mathbb{F}}\|{A^{(r)}\otimes \hat{H}^{(r)}}\|_2 \Big),
\end{align}
where $\|\cdot\|_2$ is the spectral norm.
To calculate the spectral norm, we consider
the Pauli decomposition of $H^{(r)}$ and $H^{(0)}$:
\begin{align}
\hat{H}^{(0)}= \sum_{l_0=1}^{L_0} a_{l_0} P_{l_0}, \quad \hat{H}^{(r)}= \sum_{l_r=1}^{L_r}|a_{l_r}|e^{i\phi_{l_r}} P_{l_r}, \quad \hat{H}^{(-r)} = \hat{H}^{(r)\dagger},
\end{align}
where $\{P_{l_j}\}$ are Pauli strings, $\{a_{l_0}\}$ are the coefficients of the zeroth Fourier mode of $\hat{H}(t)$, and $|a_{l_r}|e^{i\phi_{l_r}}$ are the coefficients of the Fourier mode $\hat H^{(r)}$.
With these decompositions, we then obtain
\begin{align}
    \|I\otimes \hat{H}^{(0)}\|_2 &\leq \sum_{l_0=1}^{L_0} |a_{l_0}| \|P_{l_0}\|_2=\|a_0\|_1, \\
  \sum_{r\in \mathbb{F}}\|{A^{(r)}\otimes \hat H^{(r)}}\|_2  &\leq \sum_{r\in\mathbb{F}^+}\sum_{l_r=1}^{L_r}\|(A^{(r)}e^{i\phi_{l_r}}+A^{(-r)}e^{-i\phi_{l_r}})\|_2 |a_{l_r}| \|P_{l_r}\|_2 \leq 2\sum_{r\in\mathbb{F}^+} \|a_r\|_1,
\end{align}
where $a_0$ is a vector of the coefficients $a_{l_0}$, and similarly for $a_r$, while $\|\cdot\|_1$ is the $\ell_1$ norm (one-norm of the coefficient of Pauli strings of an operator).
Here, we used eigenvalue properties of symmetrized Toeplitz matrices. Utilizing the results from Ref.~\cite{Ekstrom_Eigvals_Toeplitz}, we can express the eigenvalues of $A^{(r)}e^{i\phi_{l_r}}+A^{(-r)}e^{-i\phi_{l_r}}$ as $2\cos(\theta_r)$, where $\theta_r$ depends on $r$ and the dimension of the auxiliary space, $2^{n_a}$. Therefore $\|(A^{(r)}e^{i\phi_{l_r}}+A^{(-r)}e^{-i\phi_{l_r}})\|_2 \leq 2$.
For more details about eigenvalues of symmetrized sparse Toeplitz matrices, see Ref.~\cite{Ekstrom_Eigvals_Toeplitz}.
If there are $R$ positive Fourier modes, then we have 
\begin{align}
    \mathcal{W}\leq 4(R+\tfrac{1}{2}) \|a_{\textrm{max}}\|_1,
\end{align}
where $\|a_{\textrm{max}}\|_1= \textrm{max}\{\|a_r\|_1\}_{r\in\mathbb{F}^+ \bigcup \{0\}} $.

Since $n_a$ should satisfy Eq.~\eqref{Eq:Truncation_na}, we can choose an upper bound on $\mathcal{W}$ to calculate $n_a$ as follows:
\begin{align}
    n_a\geq \Big\lceil\log_2\Big(\frac{4(R+\frac{1}{2})\|a_{\textrm{max}}\|_1}{\Omega}\Big)\Big\rceil.
\end{align}
If we consider a time-periodic Hamiltonian with single Fourier  mode  ($R=1$) that has a number of Pauli strings that scales as $O(L^K)$, and $J_{\textrm{max}}$ the largest local interaction strength, then we can choose $n_a$  as: 
\begin{align}
    n_a\geq  \Big\lceil \log_2\Big(\frac{6L^KJ_{\textrm{max}}}{\Omega}\Big)\Big\rceil= \Big\lceil K \log_2L +\log_2\Big(\frac{6J_{\textrm{max}}}{\Omega}\Big)\Big\rceil,
\end{align}
where we have replaced the $\|a_{\textrm{max}}\|_1= L^K J_{\textrm{max}}$. 

   \section{Measurement cost scaling  of  VQE  cost function }\label{Appen:Meaurement_Scaling_of_cost_fuction}
 In this appendix, we discuss the measurement cost of Floquet-ADAPT-VQE associated with the cost function in Eq.~\eqref{Eq:Cost_function} of the main text. To this end, we analyze the number of Pauli strings and commuting sets in the extended Floquet Hamiltonian and its square, separately for the auxiliary and physical parts.


\subsection{Explicit expression of operator in the cost function}\label{Appen_Measurement_Sub1}
Here our goal is to provide the explicit expression of auxiliary part of operators in the cost functions, defined in Eq.~\eqref{Eq:Cost_function}, in terms of  $\tilde{A}^{(r)}_{\gamma_r}$ and $\{ A^\mathrm{d}, \tilde{A}^{(r)}_{\gamma_r}\}$. 
The two terms of the operators in the cost function are:
\begin{align} \label{Eq:HF_and_HFSqr}
    \hhat{H}_F &= \Omega A^\mathrm{d} +\sum_{\gamma_0=1}^{\Gamma_0} P_{\gamma_0} + \sum_{r\in \mathbb{F}^+} \sum_{\gamma_r=1}^{\Gamma_r}\tilde{A}^{(r)}_{\gamma_r}\otimes P_{\gamma_r} \nonumber\\
     \hhat{H}_F^2 &= \Omega^2 (A^{\mathrm{d}})^2 + \sum_{\gamma_0,\gamma'_0=1}^{\Gamma_0} P_{\gamma_0} P_{\gamma'_0} + \sum_{\gamma_0=1}^{\Gamma_0}2\Omega A^{\mathrm{d}} \otimes P_{\gamma_0} + \Omega \sum_{r\in \mathbb{F}^+}\sum_{\gamma_r=1}^{\Gamma_r} \{A^{\mathrm{d}},\tilde{A}^{(r)}_{\gamma_r}\}\otimes P_{\gamma_r} +  \nonumber\\
     &\hspace{1in}
     \sum_{r\in \mathbb{F}^+}\sum_{\gamma_0=1}^{\Gamma_0} \sum_{\gamma_r=1}^{\Gamma_r} \tilde{A}^{(r)}_{\gamma_r}\otimes \{P_{\gamma_0}, P_{\gamma_{r'}}\} +  \sum_{r,r'\in \mathbb{F}^+}\sum_{\gamma_r=1}^{\Gamma_r}\sum_{\gamma'_{r'}=1}^{\Gamma_{r'}} \Big( \tilde{A}^{(r)}_{\gamma_r}\tilde{A}^{(r')}_{\gamma_{r'}}\Big)\otimes P_{\gamma_{r}} P_{\gamma'_{r'}},
\end{align}
where $\tilde{A}^{(r)}_{\gamma_r}= |a_{\gamma_r}|\left(A^{(r)}e^{i\phi_{\gamma_r}}+A^{(-r)}e^{-i\phi_{\gamma_r}}\right)$ and $\{O_1,O_2\}= O_1O_2+ O_2O_1$ for operators $O_1,O_2$. We note that the Pauli strings representing $A^{(r)}$, $A^{(-r)}$, and $\tilde{A}^{(r)}$ are the same; only their coefficients change. Therefore, in the discussion related to Pauli strings in this section, we use $A^{(r)}$ and $\tilde{A}^{(r)}_{\gamma_r}=\tilde{A}^{(r)}$ interchangeably. Additionally, the notation $\gamma'_{r'}$, $\gamma_{r'}$, $\gamma'_r$, and $\gamma_r$ emphasizes that $r$ and $r'$, as well as $\gamma$ and $\gamma'$ are independent.

 Before proceeding, we emphasize that there are two ways to represent  $\hhat{H}_F^{2}$ in the truncated Floquet Hilbert space. The first approach is to truncate the auxiliary space first and then square the truncated Hamiltonian to obtain $\hhat{H}_F^{2}$. The second approach is to first square the extended Floquet Hamiltonian in the infinite-dimensional auxiliary space, make use of the algebraic properties of the auxiliary matrices, and only then apply the truncation to $\hhat{H}_{F}^{2}$.  When the auxiliary-space truncation is chosen as discussed in Appendix~\ref{Appen:truncation}, both approaches are the same. We use the second approach here because it  provides an analytical form of $\hhat{H}_F^2$ due to the relations among the auxiliary matrices, in infinite dimensional space, as:
\begin{align}
A^{(r)}A^{(r')} = A^{(r+r')}, \quad
A^{(r)}A^{(-r')} = A^{(r-r')}, \quad A^{(r)}A^{(-r)}=A^{(-r)}A^{(r)} = I,\quad
[A^{\mathrm{d}},A^{(\pm r)}] = \pm rA^{(r)} .
\end{align}
Now we focus on $\hhat{H}_F^2$. Except for the last term, all other terms in $\hhat{H}_F^2$, given in Eq.~\eqref{Eq:HF_and_HFSqr}, contain the auxiliary operators $A^\mathrm{d}$, $\{A^\mathrm{d},\tilde{A}^{(r)}\}$, or $\tilde{A}^{(r)}$. 
We express the last term of $\hhat{H}_F^2$ in Eq.~(\ref{Eq:HF_and_HFSqr}) as:
\begin{align}\label{Eq:Off-Diag_HF_partial_terms}
    \hhat{H}_{Fr,r'} &= \sum_{r,r'\in \mathbb{F}^+}\sum_{\gamma_{r}=1}^{\Gamma_{r}}\sum_{\gamma'_{r'}=1}^{\Gamma_{r'}}\tilde{A}^{(r)}_{\gamma_r} \tilde{A}^{(r')}_{\gamma_{r'}} \otimes P_{\gamma_r}P_{\gamma'_{r'}} \nonumber\\
    &= \sum_{r\in \mathbb{F}^+}\sum_{\gamma_{r}=1}^{\Gamma_{r}}\sum_{\gamma'_{r}=1}^{\Gamma_{r}}\tilde{A}_{\gamma_{r}}^{(r)} \tilde{A}_{\gamma'_{r}}^{(r)} \otimes P_{\gamma_r}P_{\gamma'_{r}} +\sum_{r\neq r'\in \mathbb{F}^+}\sum_{\gamma_{r}=1}^{\Gamma_{r}}\sum_{\gamma_{r'}=1}^{\Gamma_{r'}}\tilde{A}_{\gamma_{r}}^{(r)} \tilde{A}_{\gamma_{r'}}^{(r')} \otimes P_{\gamma_r}P_{\gamma_{r'}} \nonumber\\
    &=\sum_{r\in \mathbb{F}^+}\sum_{\gamma_{r}=1}^{\Gamma_{r}}\sum_{\gamma'_{r}=1}^{\Gamma_{r}} \Big(a_{\gamma_r}a_{\gamma'_r}A^{(2r)} +a^{*}_{\gamma_r}a^{*}_{\gamma'_r}A^{(-2r)} + a_{\gamma_r}a^{*}_{\gamma'_r}A^{(r)}A^{(-r)}+a^{*}_{\gamma_r}a_{\gamma'_r}A^{(-r)}A^{(r)} \Big) \otimes \nonumber\\ 
    &P_{\gamma_r}P_{\gamma'_{r}} +\sum_{r\neq r'\in \mathbb{F}^+}\sum_{\gamma_{r}=1}^{\Gamma_{r}}\sum_{\gamma_{r'}=1}^{\Gamma_{r'}} \Big(a_{\gamma_r}a_{\gamma_{r'}}A^{(r+r')} +a^{*}_{\gamma_r}a^{*}_{\gamma_{r'}}A^{(-r-r')}+ a_{\gamma_r}a^{*}_{\gamma_{r'}}A^{(r)}A^{(-r')}+ \nonumber\\
    &\hspace{3.5in} a^{*}_{\gamma_r}a_{\gamma_{r'}}A^{(-r)}A^{(r')} \Big) \otimes P_{\gamma_r}P_{\gamma_{r'}}   \nonumber\\ 
&=\sum_{r\in \mathbb{F}^+}\sum_{\gamma_{r}=1}^{\Gamma_{r}}\sum_{\gamma'_{r}=1}^{\Gamma_{r}} \Big(\tilde{A}^{(2r)}(a_{\gamma_r},a_{\gamma'_r})+ 2\textrm{Re}\{a_{\gamma_r}a^{*}_{\gamma'_r}\}I\Big)\otimes P_{\gamma_r}P_{\gamma'_{r}} + \sum_{r\neq r'\in \mathbb{F}^+}\sum_{\gamma_{r}=1}^{\Gamma_{r}}\sum_{\gamma_{r'}=1}^{\Gamma_{r'}} \nonumber\\
&\hspace{1.7in}\Big(\tilde{A}^{(r+r')}(a_{\gamma_r}a_{\gamma_{r'}})+  \tilde{A}^{(|r-r'|)}(a_{\gamma_r}a^{*}_{\gamma_{r'}})\Big) \otimes P_{\gamma_r}P_{\gamma_{r'}}, 
\end{align}
where we have denoted the Hermitian auxiliary matrices, in the last line, as $\tilde{A}^{(k)}(\alpha)=  A^{(k)}(\alpha)+ A^{(-k)}(\alpha^{*})$ for complex number $\alpha$. From Eqs.~\eqref{Eq:HF_and_HFSqr} and \eqref{Eq:Off-Diag_HF_partial_terms}, we see that the auxiliary parts of the operators appearing in the cost function can be represented as $\tilde{A}^{(r)}_{\gamma_r}$ and $\{ A^\mathrm{d}, \tilde{A}^{(r)}_{\gamma_r}\}$.

We note that if the number of positive Fourier modes is $R$, then the total number of terms in the summation in the Eq.~\eqref{Eq:HF_and_HFSqr}, corresponding to auxiliary  and physical parts, is $O(R^2)$. Each term in the summation contains either $A^\mathrm{d}, \{A^\mathrm{d}, \tilde{A}^{(r)}\}$ or $ \tilde{A}^{(r)}$ in the auxiliary part, and $ \{P_\gamma\}_{\gamma}, \{P_\gamma P_{\gamma'}\}_{\gamma,\gamma'}$ in the physical part. Let $f(n_a)$ represents the number of commuting sets corresponding to auxiliary part in general case such that it could be either $O(\exp(n_a))$ or $O(\mathrm{poly}(n_a))$. In general, the total number of commuting sets scales as $O\left(R^2 f(n_a)\,\mathrm{poly}(L)\right)$, where the number of commuting partitions for the time-periodic Hamiltonian scales as  $O\left(\mathrm{poly}(L)\right)$.

Now we consider a time-periodic Hamiltonian having geometrically local Pauli strings of maximum weight $w$. The Pauli strings corresponding to the physical part of $\hhat{H}_F$ form a subset of the Pauli strings representing $H(0)=H^{(0)}+\sum_{r\in \mathrm{\mathbb{F}}} H^{(r)}$. The following lemma bounds the number of partitions required for these Pauli strings in the physical part of the extended Floquet Hamiltonian and its square. Furthermore, in the lemma, we consider $H \equiv H(0)$.

\textbf{Lemma 1}
A Hamiltonian $H$ composed of geometrically local Pauli strings with maximum weight $w$ can be partitioned into $O(w)$ commuting sets. The number of commuting partitions required for the Pauli strings in $H^2$ scales as $O(L)$.

\textbf{Proof:} Let $H=\sum_{\gamma=1}^{\Gamma}\beta_\gamma P_{\gamma}$ ,
where each $P_\gamma$ is a geometrically local Pauli string of weight at most $w$ and $\Gamma\in O(L)$. Then we have
\[
H^2=\sum_{\gamma=1}^{\Gamma}\sum_{\gamma'=1}^{\Gamma} \beta_\gamma \beta_{\gamma'} P_\gamma P_{\gamma'} .
\]
Equivalently, the terms contributing to $H^2$ are products of the form $P_\gamma P_{\gamma'}$. For each fixed value of $\gamma$, the strings $P_\gamma P_{\gamma'} $ inherit the locality structure from $P_{\gamma'} $ (apart from the multiplication of $P_\gamma$ overall). Therefore, for each fixed $\gamma$, these terms can again be partitioned into $O(w)$ commuting sets. Since the maximum possible value of $\gamma$ is $\Gamma$, the total number of commuting sets needed for $H^2$ is at most $O(w\Gamma)= O(wL)$.
In particular, when $w=O(1)$, the number of commuting sets scales as
$O(L)$. $\square$

We now determine the commuting sets of Pauli strings in $\hat{\hat{H}}_F^2$ for a time-periodic Hamiltonian whose terms are geometrically local Pauli strings of maximum weight $w$. In the worst case, the auxiliary part can be represented using $4^{n_a}$, or $O\left((L J{\max}/\Omega)^2\right)$, Pauli strings. On the physical part, we have $O(L)$ commuting sets of Pauli strings. Thus, we obtain $O(L^3)$ commuting sets of Pauli strings corresponding to $\hat{\hat{H}}_F^2$ for a time-periodic Hamiltonian involving only weighted geometrically local Pauli strings.

In many physical settings, the Hamiltonian contains only a single Fourier mode. For example, this occurs when a material is driven by a monochromatic laser field. In this case, the number of commuting sets corresponding to the auxiliary part of $\hhat{H}_F^2$ scales polynomially with $n_a$, or polylogarithmically with $L$, as shown in the next subsection.


\subsubsection{Number of commuting set of Pauli strings corresponding to auxiliary matrices}
Since we need to consider the odd dimensional auxiliary space, we consider the symmetric auxiliary matrix as $ A_{\text{sym}}^{(1)}=  A^{(1)}- A_{\text{asy}}^{(1)}$. Let $S_{n_a}^{\mathrm{d}}$, $S_{n_a}^{(r)}$, $S_{n_a,\mathrm{asy}}^{(r)}$, and $S_{n_a,\mathrm{sym}}^{(r)}$ denote the sets of Pauli strings corresponding to $A^{\mathrm{d}}$, $A^{(r)}$, $A_{\mathrm{asy}}^{(r)}$, and $A_{\mathrm{sym}}^{(r)}$, respectively, for an $n_a$-qubit truncation. Our goal is to partition $S_{n_a,\mathrm{sym}}^{(r)}$ into mutually commuting subsets of Pauli strings.
We start with explicit counting of the minimum number of commuting subsets of  Pauli strings for $r=1$ and $n_a=2,3$. For $n_a=2$, we have: 
\begin{align}
S^{\mathrm{d}}_2&= \{Z_1,Z_2\}, \ S^{(1)}_2= \{ XX,XY,YX,YY,IX,IY\}, \ S^{(1)}_{2,\textrm{asy}}= \{IX,IY,ZX,ZY\}, \\
S^{(1)}_{2,\textrm{sym}}&= \big\{ XX,YY\big\} \bigcup \{XY,YX\big\}\bigcup\big\{IX,ZX\big\}\bigcup\big\{IY, ZY\big\}.
\end{align}
So, number of commuting set is $4$ for $n_a=2$.

For $n_a=3$ we have: 
\begin{align}
  S^{(1)}_3&= \{ XXX,XYX,XXY,XYY,YXX,YXY,YYX,YYY, IXX,IXY,IYX,IYY,IIX,IIY\}, \nonumber\\
S^{\mathrm{d}}_3&= \{Z_1,Z_2,Z_3\},\ S^{(1)}_{3\textrm{asy}}= \{IIX,IIY,IZX,IZY,ZIX,ZIY,ZZX,ZZY\}, \nonumber\\
S^{(1)}_{3,\textrm{sym}}&= \big\{ IXX,IYY,XXX,XYY\big\} \bigcup \big\{ IXY,IYX,XXY,XYX\big\} \bigcup \{YYY, YXX, XXY,XYX\big\} \nonumber\\
&\hspace{1in} \bigcup\big\{ZZX,ZIX,IZX,IIX\big\}\bigcup\big\{ZZY,ZIY,IZY,IIY\big\}. 
\end{align}
Thus, number of commuting set is $5$ for $n_a=3$.


\textbf{Theorem B1.}
The set of Pauli strings required to represent $A_{\mathrm{sym}}^{(1)}$ in an $n_a$-qubit auxiliary space can be partitioned into at most $n_a+2$ mutually commuting subsets.

\textbf{Proof.}
Let $S_{\mathrm{sym}}^{(1)}$, $S_{n_a}^{(1)}$, and $S_{\mathrm{asy}}^{(1)}$ denote the sets of Pauli strings appearing in the decompositions of $A_{\mathrm{sym}}^{(1)}$, $A^{(1)}$, and $A_{\mathrm{asy}}^{(1)}$, respectively in $n_a$-qubit auxiliary space. Since $ A_{\mathrm{sym}}^{(1)}=A^{(1)}-A_{\mathrm{asy}}^{(1)},
$ we have
$$
S_{\mathrm{sym}}^{(1)}\subseteq S_{n_a}^{(1)}\cup S_{\mathrm{asy}}^{(1)}.
$$
Thus, it is sufficient to construct a commuting partition of $S_{n_a}^{(1)}\cup S_{\mathrm{asy}}^{(1)}$.

We first consider $S_{n_a}^{(1)}$. From the Pauli decomposition of $A^{(1)}$, this set can be written as
$$ S_{n_a}^{(1)}=\bigcup_{k=1}^{n_a} S_{n_a,k}^{(1)},
\qquad
S_{n_a,k}^{(1)}=\left\{ I^{\otimes(n_a-k)}\otimes \widetilde{Q}_k:
\widetilde{Q}_k\in\{X,Y\}^{\otimes k}
\right\}.
$$
Thus, each element of $S_{n_a,k}^{(1)}$ has nontrivial support only on the $k$ rightmost qubits. For two Pauli strings in $\{X,Y\}^{\otimes k}$, the commutation relation is determined by the parity of the number of sites at which the two strings differ: they commute if and only if this number is even. The Pauli strings in $S_{n_a}^{(1)}$ can therefore be grouped according to these parity constraints. Since $S_{n_a,k}^{(1)}$ requires two commuting groups, and $S_{n_a,k+1}^{(1)}$ also admits a partition into two commuting groups, one might expect four groups in total. However, one commuting group of $S_{n_a,k+1}^{(1)}$ can be merged with a group of $S_{n_a,k}^{(1)}$, because the $k$ rightmost Pauli operators are identical and the
strings differ only at one additional position. Hence, each increase in $k$ adds only one new commuting group. For $k=2,\ldots,n_a$, one can construct $n_a$ mutually commuting subsets that contain all strings in $S_{n_a}^{(1)}\setminus S_{n_a,1}^{(1)}$. 

The two strings in $S_{n_a,1}^{(1)}$, namely $I^{\otimes(n_a-1)}X,     I^{\otimes(n_a-1)}Y$, will be included in the two commuting subsets associated with $S_{\mathrm{asy}}^{(1)}$, as described below. Now consider $S_{\mathrm{asy}}^{(1)}$. Every Pauli string in this set has either $X$ or $Y$ on the rightmost qubit, while all other qubits contain only $I$ or $Z$. Therefore, $S_{\mathrm{asy}}^{(1)} = S_{\mathrm{asy},X}^{(1)} \cup S_{\mathrm{asy},Y}^{(1)},$
where
$S_{\mathrm{asy},X}^{(1)}=\left\{ R\otimes X: R\in\{I,Z\}^{\otimes(n_a-1)}\right\},$ and
$ S_{\mathrm{asy},Y}^{(1)} =\left\{R\otimes Y: R\in\{I,Z\}^{\otimes(n_a-1)} \right\}$. Each of these two sets is mutually commuting, because the only nontrivial $X$ or $Y$ operator appears on the same rightmost qubit, while all remaining operators are diagonal. Moreover, these two sets contain the strings $I^{\otimes(n_a-1)}X$, $ I^{\otimes(n_a-1)}Y,$ respectively.

Combining the $n_a$ commuting subsets for $S_{n_a}^{(1)}\setminus S_{n_a,1}^{(1)}$ with the two commuting subsets for $S_{\mathrm{asy}}^{(1)}$, we obtain a partition of $S_{n_a}^{(1)}\cup S_{\mathrm{asy}}^{(1)}$ into at most $n_a+2$  commuting subsets. Since $S_{\mathrm{sym}}^{(1)}\subseteq S_{n_a}^{(1)}\cup S_{\mathrm{asy}}^{(1)},$ the same upper bound applies to $S_{\mathrm{sym}}^{(1)}$.
\(\square\)

\textbf{Corollary B1.}
For an $n_a$-qubit auxiliary space, the Pauli strings representing $A_{\mathrm{sym}}^{(r=2^{k_0})}$, where $k_0$ is a non-negative integer, can be partitioned into at most $n_a-k_0+2$ commuting subsets.

\textbf{Proof:} 
 From Eq.~\eqref{Eq:Aux_Mat_r} and the discussion in Appendix~\ref{Appen:Cal_Paui_Off}, we have for $k_0\in\mathbb{Z}^+$,
\begin{equation}
A_{n_a}^{(r=2^{k_0})}=\sum_{\ell=1}^{n_a-k_0} I^{\otimes(n_a-k_0-\ell)} \otimes(P^-)\otimes(P^+)^{\otimes \ell}\otimes I^{\otimes k_0}.
\end{equation}
Let a set of Pauli strings representing $A_{n_a}^{(r=2^{k_0})}$ be
\begin{equation}
S_{n_a}^{(r=2^{k_0})} =\bigcup_{\ell=1}^{n_a-k_0} I^{\otimes(n_a-k_0-\ell)} \otimes \widetilde{Q}_{\ell}\otimes I^{\otimes k_0},
\end{equation}
where $\widetilde{Q}_{\ell}\in \{X,Y\}^{\otimes \ell}$. For $A_{n_a,\mathrm{asy}}^{(r=2^{k_0})}$, the set of Pauli strings will have   $\{X,Y\}$ only at a single position, while all remaining non-identity operators are $Z$. The auxiliary matrices corresponding to the $r=2^0=1$ case and the $r=2^{k_0}$ case are similar; the only difference is that the Pauli weight is reduced by $k_0$ in the latter case. Therefore, we can follow the same argument and obtain $n_a-k_0+2$ commuting sets for $S_{n_a,\mathrm{sym}}^{(r=2^{k_0})}$.
$\square$

\textbf{Theorem B2.}
The Pauli strings corresponding to the operators $\{A^{\mathrm d},A_{\mathrm{sym}}^{(1)}\}$ in an $n_a$-qubit auxiliary space can be partitioned into $O(n_a^2)$ commuting sets.

\textbf{Proof.}
To prove this, we work with $\{A^\mathrm{d},A_{\mathrm{sym}}^{(1)}\}$ directly. Since $A^{\mathrm d}=I+\sum_{j=1}^{n_a}d_j Z_j$, we have
\begin{equation}
\{A^{\mathrm d},A_{\mathrm{sym}}^{(1)}\}
= 2A_{\mathrm{sym}}^{(1)} + \sum_{j=1}^{n_a}d_j\{Z_j,A_{\mathrm{sym}}^{(1)}\}.
\label{eq:anticom_Adsym}
\end{equation}
The first term, $2A_{\mathrm{sym}}^{(1)}$, contains the same Pauli strings as $A_{\mathrm{sym}}^{(1)}$. From the Pauli decomposition of $A_{\mathrm{sym}}^{(1)}$, these strings can already be organized into $n_a+2$ commuting sets as shown in Theorem B1. We can write $\{Z_j,A_{\mathrm{sym}}^{(1)}\}= \{Z_j,A^{(1)}\}-\{Z_j,A_{\mathrm{asy}}^{(1)}\}$. Let $\{S_{n_a,k}^{(1)}\}_{k=1}^{n_a}$ denote the sets of Pauli strings, with weights up to $n_a$, corresponding to $A^{(1)}$. Each $Z_l$ commutes only with the sets $\{S_{n_a,k}^{(1)}\}_{k=1}^{l}$ for $l=\{1,2,\ldots,n_a-1\}$. Therefore, the anti-commutator $\{\sum_{j=1}^{n_a} Z_j,A^{(1)}\}$ contains contributions from $n_a(n_a-1)/2$ such terms. The Pauli strings appearing in $\{Z_j,A^{(1)}\}$ differ from those in $A^{(1)}$ only by replacing some identity operators $I$ with $Z$ operators. Since each contribution can be partitioned into two commuting sets, the total number of commuting sets scales as $O(n_a^2)$.

Next, we consider the terms arising from $\{A^{\mathrm d},A_{\mathrm{asy}}^{(1)}\}$. Since the Pauli strings representing $A_{\mathrm{asy}}^{(1)}$ only anticommute with $Z_1$, the resulting terms consist of $\{I,A_{\mathrm{asy}}^{(1)}\}$ and $\{Z_j,A_{\mathrm{asy}}^{(1)}\}$ for $j=2,\ldots,n_a$. All these terms contain either $X$ or $Y$ on the rightmost qubit and only $I$ or $Z$ on the remaining qubits. Thus, there are only two commuting sets for $\{A^\mathrm{d},A_{\mathrm{asy}}^{(1)}\}$. Therefore, the total number of commuting sets associated with $\{A^{\mathrm d},A_{\mathrm{sym}}^{(1)}\}$ scales as $O(n_a^2)$. \(\square\)


From Theorems B1 and B2 and Corollary B1, we see that, for the single-Fourier-mode case, the auxiliary-space parts of $\hhat{H}_F^2$ can be decomposed into $O(n_a^2)$ commuting sets of Pauli strings. Hence, if the physical part of $\hhat{H}_F^2$ contains $O(L^{2k})$ Pauli strings, the full auxiliary--physical decomposition contains at most $O(n_a^2 L^{2k})$ Pauli strings. 


Finally, utilizing the discussion in this section, we obtain the following Theorem for a particular class of periodically driven Hamiltonians.

\textbf{Theorem  B3 }  For a time-periodic Hamiltonian with a single Fourier mode ($|r|=1$) and $w$-weighted geometrically local Pauli strings, the number of commuting terms scales as
$O(n_a^2 L)=O\!\left((\log^2 L)L\right)$.

\textbf{Proof:}
Since we have $|r|=1$, the squared extended Floquet Hamiltonian $\hhat{H}_F^2$
contains terms of the form  $A^\mathrm{d}$, $\tilde{A}_{\mathrm{sym}}^{(1)}, \tilde{A}^{(2)}_{\mathrm{sym}}$, and $\{A^\mathrm{d},A_{\mathrm{sym}}^{(1)}\}$ acting on the auxiliary part, together with physical Pauli strings $\sum_{\ell,\ell'} P_\ell P_{\ell'}$ . The auxiliary part and the physical part contain $O(n_a^2)$ and $O(L)$ commuting Pauli strings, respectively. Therefore, the total number of commuting terms scales as $O(n_a^2 L)$.
Using $n_a=O(\log L)$, this becomes $O(n_a^2 L)=O\left((\log^2 L)L\right)$.
$\square$

\section{Choice of initial auxiliary state}\label{Appen:initial_state}
In this appendix, we prove that the choice of the initial state $\kket{\psi_\mathrm{in}}=\ket{0}\otimes\ket{\phi_\mathrm{p}}$ as given in Eq.~(\ref{Eq:init_state}) eliminates  the dependency of its squared Floquet energy on driving frequency $\Omega$ and on the number of auxiliary qubits $n_a$.  We assume that $N_c > \max_{r\in\mathbb{F}} |r|$ in the following derivation.
We begin by writing $\hhat{H}_\mathrm{F}^2$ in terms of its dependence on $\Omega$:
\begin{flalign}
 \hhat{H}_\mathrm{F}^2 &=I \otimes (\hat H^{(0)})^2 + \hhat{g}_{\Omega} + \hhat{h}, \\
 &\hspace{-48mm}\text{where}  \nonumber\\
\hhat{g}_{\Omega} &= \Omega^2\big(A^{\mathrm{d}})^2\otimes  \hat I + 2\Omega ( A^{\mathrm{d}}\otimes\hat H^{(0)}) + \Omega\sum_{r\in \mathbb{F}}\big\{A^{\mathrm{d}},A^{(r)}_\mathrm{sym}\big\}\otimes \hat H^{(r)}, \nonumber\\
 \hhat{h}&= \sum_{r\in \mathbb{F}}A^{(r)}_\mathrm{sym}\otimes\big\{\hat H^{(0)},\hat H^{(r)}\big\}+  \Big(\sum_{r\in \mathbb{F}}A^{(r)}_\mathrm{sym}\otimes \hat H^{(r)}\Big)^2. \nonumber
\end{flalign}
Here, $\{O_1,O_2\}=O_1O_2+O_2O_1$ is the anti-commutator, and $A^{(r)}_\mathrm{sym}=A^{(r)}-A^{(r)}_\mathrm{asy}$. 
We choose $\kket{\psi_\mathrm{in}}= \ket{0}\otimes\ket{\phi_\mathrm{p}}$. Let us first look at the terms that depend on $\Omega$:
\begin{align}
\bbra{\psi_\mathrm{in}}\hhat{g}_{\Omega}\kket{\psi_\mathrm{in}} &= \bbra{\psi_\mathrm{in}}\Omega^2\big(A^{\mathrm{d}})^2\otimes  \hat I \kket{\psi_\mathrm{in}} + 2\Omega \bbra{\psi_\mathrm{in}} (A^{\mathrm{d}}\otimes\hat H^{(0)})\kket{\psi_\mathrm{in}}  + \nonumber \\
&\hspace{2in}\bbra{\psi_\mathrm{in}}\Omega\sum_{r\in \mathbb{F}}\big\{A^{\mathrm{d}},A^{(r)}_\mathrm{sym}\big\}\otimes \hat H^{(r)} \kket{\psi_\mathrm{in}}\nonumber\\
    &= \Omega^2\bra{0}\big(A^{\mathrm{d}})^2\ket{0}  \bra{\phi_\mathrm{p}}\hat I \ket{\phi_\mathrm{p}} + 2\Omega\bra{0} A^{\mathrm{d}}\ket{0} \bra{\phi_{\mathrm{p}}}\hat{H}^{(0)} \ket{\phi_{\mathrm{p}}}  +
    \nonumber \\
&\hspace{2in}\Omega\sum_{r\in \mathbb{F}}\bra{0}\big\{A^{\mathrm{d}},A^{(r)}_\mathrm{sym}\big\} \ket{0}\bra{\phi_\mathrm{p}}\hat{H}^{(r)}\ket{\phi_\mathrm{p}} \nonumber\\ 
    &=0.
\end{align}
Thus this choice of the initial state removes the dependence on $\Omega$. Now we look at the terms of $\hhat{H}_\mathrm{F}^2$ that do not contain $\Omega$ but still depend on $n_a$: 
\begin{align}
    \bbra{\psi_\mathrm{in}}I \otimes (\hat H^{(0)})^2\kket{\psi_\mathrm{in}} 
    &= \sandwich{0}{I}{0} \sandwich{\phi_\mathrm{p}}{(\hat H^{(0)})^2}{\phi_\mathrm{p}} \nonumber\\ 
    &=\sandwich{\phi_\mathrm{p}}{(\hat H^{(0)})^2}{\phi_\mathrm{p}}.
    \end{align}
    Thus, the above expectation value is independent of $n_a$. We also have
\begin{align}\label{Eq:Init_State}
    \bbra{\psi_\mathrm{in}}\hhat{h}\kket{\psi_\mathrm{in}}&= \bbra{\psi_\mathrm{in}}\sum_{r\in \mathbb{F}}A^{(r)}_\mathrm{sym}\otimes\big\{\hat H^{(0)},\hat H^{(r)}\big\} \kket{\psi_\mathrm{in}} + \bbra{\psi_\mathrm{in}}\Big(\sum_{r\in \mathbb{F}}A^{(r)}_\mathrm{sym}\otimes \hat H^{(r)}\Big)^2 \kket{\psi_\mathrm{in}} \nonumber\\
    &= \sum_{r\in \mathbb{F}}\bra{0}A^{(r)}_\mathrm{sym}\ket{0}  \bra{\phi_\mathrm{p}} \big\{\hat H^{(0)},\hat H^{(r)}\big\}\ket{\phi_\mathrm{p}} + \sum_{r,r'\in \mathbb{F}}\bra{0}A^{(r)}_\mathrm{sym}A^{(r')}_\mathrm{sym} \ket{0}\bra{\phi_\mathrm{p}}\hat{H}^{(r)}\hat{H}^{(r')}\ket{\phi_\mathrm{p}} \nonumber \\
   &=\sum_{r\in \mathbb{F}}c_1(r)  \bra{\phi_\mathrm{p}} \big\{\hat H^{(0)},\hat H^{(r)}\big\}\ket{\phi_\mathrm{p}} + \sum_{r,r'\in \mathbb{F}}c_2(r,r')\bra{\phi_\mathrm{p}}\hat{H}^{(r)}\hat{H}^{(r')}\ket{\phi_\mathrm{p}}, 
\end{align}
where $c_1(r)$ and $  c_2(r,r')$ are dependent on Fourier indices but not on $n_a$. Let's look at $c_1(r)$
\begin{align}
    c_1(r>0)&= \sum_{n=-N_\mathrm{c}}^{N_\mathrm{c}-r}\bra{0}\bigl(\ket{n+r}\bra{n}\bigr)\ket{0} =0, \nonumber\\
     c_1(r<0)&= \sum_{n=-N_\mathrm{c}}^{N_\mathrm{c}-|r|}\bra{0}\bigl(\ket{n}\bra{n+|r|}\bigr)\ket{0} =0.
\end{align}
Now, we look at the $c_2(r,r')$ corresponding to four different values of $(r,r')$
\begin{align}
    c_2(r>0,r'>0)&= \bra{0}\Bigl(\sum_{n=-N_\mathrm{c}}^{N_\mathrm{c}-r}\ket{n+r}\bra{n}\Bigr)\Bigl(\sum_{m=-N_\mathrm{c}}^{N_\mathrm{c}-r'}\ket{m+r'}\bra{m}\Bigr)\ket{0} \nonumber\\
    &= \Bigl(\sum_{n=-N_\mathrm{c}}^{N_\mathrm{c}-r}\delta_{0,n+r}\bra{n}\Bigr)\Bigl(\sum_{m=-N_\mathrm{c}}^{N_\mathrm{c}-r'}\ket{m+r'}\delta_{m,0}\Bigr) \nonumber\\
    &= \inner{-r}{r'}=0,\nonumber\\
    c_2(r<0,r'<0)&= \bra{0}\Bigl(\sum_{n=-N_\mathrm{c}}^{N_\mathrm{c}-|r|}\ket{n}\bra{n+|r|}\Bigr)\Bigl(\sum_{m=-N_\mathrm{c}}^{N_\mathrm{c}-|r'|}\ket{m}\bra{m+|r'|}\Bigr)\ket{0} \nonumber\\
    &= \inner{(|r|)}{(-|r'|)}=0,\nonumber\\
    c_2(r>0,r'<0)&= \bra{0}\Bigl(\sum_{n=-N_\mathrm{c}}^{N_\mathrm{c}-r}\ket{n+r}\bra{n}\Bigr)\Bigl(\sum_{m=-N_\mathrm{c}}^{N_\mathrm{c}-|r'|}\ket{m}\bra{m+|r'|}\Bigr)\ket{0}  \nonumber\\
    &= \inner{-r}{(-|r'|)}=\delta_{r,|r'|}=\delta_{r,-r'}, \nonumber\\ 
    c_2(r<0,r'>0)&= \bra{0}\Bigl(\sum_{n=-N_\mathrm{c}}^{N_\mathrm{c}-|r|}\ket{n}\bra{n+|r|}\Bigr)\Bigl(\sum_{m=-N_\mathrm{c}}^{N_\mathrm{c}-r'}\ket{m+r'}\bra{m}\Bigr)\ket{0}  \nonumber\\
    &= \inner{-r}{(|r'|)}=\delta_{|r|,r'}=\delta_{-r,r'}.
\end{align}
Putting the above values of  $c_1(r)$ and $c_2(r,r')$ in Eq.~(\ref{Eq:Init_State}), we obtain
\begin{align}
\bbra{\psi_\mathrm{in}}\hhat{h}\kket{\psi_\mathrm{in}}&= 0 + \sum_{r\in\mathbb{F}^{+},r'\in\mathbb{F}^{-}}\bra{\phi_\mathrm{p}}\hat{H}^{(r)}\hat{H}^{(r')}\ket{\phi_\mathrm{p}}\delta_{r,-r'} +\sum_{r'\in\mathbb{F}^{+}, r\in\mathbb{F}^{-}}\bra{\phi_\mathrm{p}}\hat{H}^{(r)}\hat{H}^{(r')}\ket{\phi_\mathrm{p}}\delta_{-r,r'} \nonumber\\
&=\sum_{r\in\mathbb{F}^{+}}\bra{\phi_\mathrm{p}}\bigl(\hat{H}^{(r)}\hat{H}^{(-r)}+ \hat{H}^{(-r)}\hat{H}^{(r)}\bigr)\ket{\phi_\mathrm{p}}, 
\end{align}
where $\mathbb{F}^{\pm}$ represent the sets of positive ($+$) and negative ($-$) indices labeling the Fourier modes of the time-periodic Hamiltonian. Overall, the squared Floquet energy for the initial state is 
\begin{align}
\bbra{\psi_\mathrm{in}}\hhat{H}_{\mathrm{F}}^2\kket{\psi_\mathrm{in}}&=\bra{\phi_\mathrm{p}}(\hat{H}^{(0)})^2\ket{\phi_\mathrm{p}}+\sum_{r\in\mathbb{F}^{+}}\bra{\phi_\mathrm{p}}\bigl(\hat{H}^{(r)}\hat{H}^{(-r)}+\hat{H}^{(-r)}\hat{H}^{(r)}\bigr)\ket{\phi_\mathrm{p}}. 
\end{align}
Thus, our choice of initial state removes the dependence on both $\Omega$ and $n_a$ in the initial squared Floquet energy, as we claimed in Sec.~\ref{sec:Prep_Floq_state}.

\begin{figure}[H]
    \centering
    \includegraphics[width=0.7\linewidth]{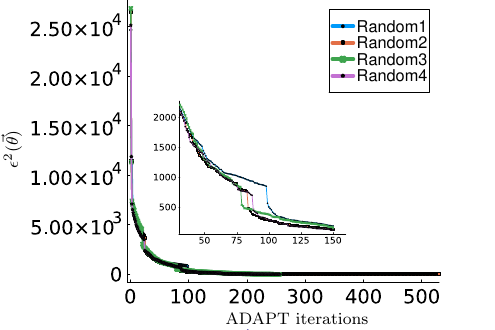}
    \caption{Squared quasienergy  $\epsilon^2(\vec{\theta})$ versus the number of ADAPT iterations for various parameters. The inset shows a zoom-in of the first 150 iterations. All parameters are expressed in units of the characteristic coupling constant $J_0$. The parameters are $L=3,n_a=4,\Omega=35.0 J_0$, $ (\bar{J}_x,\bar{J}_y,\bar{J}_z)=(3.7, 2.8, 3.9) J_0, (\delta J_x,\delta J_y,\delta J_z)=(0.0,0.0,0.0) J_0$, $\bar B_z= 2.9 J_0,\delta B_z=2.7 J_0$. We use a mixed Floquet operator pool consisting of all possible Pauli operators on the auxiliary qubits and all weight-2 Pauli strings on the physical qubits.}
    \label{Fig:Appen_Initial_State_Omg35}
\end{figure}

 Next, using numerical simulation, we look into how the squared Floquet energies of random states depend on $\Omega$. Figure~\ref{Fig:Appen_Initial_State_Omg35} shows how the squared Floquet energy changes with the number of ADAPT iterations for four random states at $\Omega= 35.0$ and $n_a = 4$. The inset shows a zoom-in over a subset of the ADAPT iterations.
  We find that the Floquet squared energy for random states is almost one order of magnitude larger than the results shown in Fig.~\ref{Fig:Initial_State}(a) of the main text for $\Omega=5.0$. Thus, we see that increasing $\Omega$ results in a higher initial squared Floquet energy, which affects the convergence rate.

\section{Expectation value of an operator} \label{App:Floq_Expt} 
In this appendix, we show how to calculate the expectation value of an operator in a Floquet state within the Floquet-Hilbert space. We provide detailed information regarding Eq. (\ref{Eq:Expt_Expression}) in the main text. For a given operator $\hat O$ and Floquet state $\ket{\psi(t)}=e^{-i\epsilon t}\ket{\phi(t)}$,  we have $\braket{\hat O(t)}_{\psi} = \sandwich{\psi(t)}{\hat O}{\psi(t)}= \sandwich{\phi(t)}{\hat O}{\phi(t)}=\braket{\hat O(t)}_{\phi}$. We can write 
\begin{align}
\braket{\hat O(t)}_{\phi} &= \sandwich{\phi(t)}{\hat O}{\phi(t)}\nonumber\\
&=\Big(\sum_{n}e^{-i n\Omega t}\bra{\phi^{(n)}}\Big)\hat O \Big(\sum_{m}e^{i m\Omega t}\ket{\phi^{(m)}} \Big) \nonumber\\
    &= \Big(\dots, \bra{\phi^{(-2)}}e^{i2\Omega t}, \bra{\phi^{(-1)}}e^{i\Omega t}, \bra{\phi^{(0)}},   \bra{\phi^{(1)}}e^{-i \Omega t}, \bra{\phi^{(2)}}e^{-i 2\Omega t},\dots \Big)
        \begin{pmatrix}
        \vdots\\
      \hat O \sum_{m}e^{i m\Omega t}\ket{\phi^{(m)}}\\
      \hat O \sum_{m}e^{i m\Omega t}\ket{\phi^{(m)}}\\
     \hat O \sum_{m}e^{i m\Omega t}\ket{\phi^{(m)}}\\
     \hat O \sum_{m}e^{i m\Omega t}\ket{\phi^{(m)}}\\
     \hat O \sum_{m}e^{i m\Omega t}\ket{\phi^{(m)}}\\
     \hat O \sum_{m}e^{i m\Omega t}\ket{\phi^{(m)}}\\
      \vdots
    \end{pmatrix}  \nonumber\\
    &= \Big(\dots,  \bra{\phi^{(-1)}}e^{i\Omega t}, \bra{\phi^{(0)}},   \bra{\phi^{(1)}}e^{-i \Omega t},  \dots \Big)
     \begin{pmatrix}
         \hat O    & \hat O  & \dots&  \hat O  &    \hat O\\
          \hat O    & \hat O  & \dots&     \hat O  &  \hat O\\
             \hat O    & \hat O  & \dots&   \hat O  &  \hat O\\
            \vdots & & &  & \vdots\\
                \hat O    & \hat O  & \dots& \hat O  &    \hat O\\
                \hat O    & \hat O  &  \dots& \hat O   &  \hat O\\
                 \hat O    & \hat O  & \dots&  \hat O  &   \hat O  \\
    \end{pmatrix}   
   \begin{pmatrix}
        \vdots \\
           e^{-i2\Omega t}\ket{\phi^{(-2)}}\\
    e^{-i\Omega t}\ket{\phi^{(-1)}}\\
     \ket{\phi^{(0)}}\\
     e^{i\Omega t} \ket{\phi^{(1)}}\\
     e^{i2\Omega t} \ket{\phi^{(2)}} \\
      \vdots
    \end{pmatrix} \nonumber
    \end{align}
    \begin{align}
    &= \Big(\dots, \bra{\phi^{(-2)}}, \bra{\phi^{(-1)}}, \bra{\phi^{(0)}},  \bra{\phi^{(1)}}, \bra{\phi^{(2)}},\dots \Big)\hhat U^{\dagger}(t)
     \begin{pmatrix}
          \hat O    & \hat O  & \dots &  \hat O  &   \hat O  \\
           \hat O    & \hat O  & \dots &  \hat O  &   \hat O  \\
             \hat O    & \hat O  & \dots & \hat O  &   \hat O  \\
            \vdots & & &   & \vdots\\
            \hat O    & \hat O  & \dots & \hat O  &   \hat O  \\
               \hat O    & \hat O  & \dots & \hat O  &   \hat O  \\
                \hat O    & \hat O  & \dots &  \hat O  &   \hat O  \\
    \end{pmatrix}   
   \hhat U(t)\begin{pmatrix}
        \vdots \\
           \ket{\phi^{(-2)}}\\
    \ket{\phi^{(-1)}}\\
     \ket{\phi^{(0)}}\\
     \ket{\phi^{(1)}}\\
     \ket{\phi^{(2)}} \\
      \vdots
    \end{pmatrix}\nonumber\\
  &=\bbra{\phi}\hhat U^{\dagger} (t)\hhat O\hhat U (t)\kket{\phi}.
\end{align}
Here, $\hhat U(t)= e^{i\Omega A^{\mathrm{d}} t}\otimes \hat I$ applies phases to the vector elements of the extended Floquet state. We note that an element in an extended Floquet state is a vector in physical space, therefore we refer to it as a vector element. Thus, we obtain Eq. (\ref{Eq:Expt_Expression}) in the main text. We have obtained this equation without truncating the auxiliary space. Next, we express the state and operators for the truncated case. Because our $\hhat{H}_F$ is block-diagonalized, the extended Floquet state in the central Floquet zone is:
\begin{align}\label{Eq:VQE_Ket_state}
    \kket{\phi}= (\ket{\phi^{(-N_{\mathrm{c}})}},\ldots,\ket{\phi^{(0)}}, \ket{\phi^{(N_{\mathrm{c}})}},0)^{\mathrm{T}},
\end{align}
where $0$ is a $D-$dimensional zero vector with $D$ being the dimension of physical Hilbert space. This implies that the truncated matrix representation of $\hhat{O}$ can have any physical operator on the last column and row without affecting the expectation values. Specifically, we choose the $\hat{O}$ operator for the last column and row. Thus, we obtain $\hhat{O} = (I+X)^{\otimes n_a} \otimes \hat{O}$. 
\section {Quasienergy spectra and eigenstates in a Floquet zone}\label{Appen: Excited_state_VQE}
In this appendix, we use an alternate method to calculate the full quasienergy spectrum within the central Floquet zone. In particular, we use a modified cost function for calculating excited states~\cite{Higgott2019variationalquantum} that penalizes overlap with the previously found eigenstates, where the cost function for the $k^{\text{th}}$ excited state may be written as
\begin{equation}\label{excited_state_cost_fn}
    C(\vec{\theta})= \bbra{\Psi(\vec{\theta})}\hhat{H}_\mathrm{F}^2\kket{\Psi(\vec{\theta})} +\sum_{i=0}^{k-1} \beta_i |\iinner{\Psi(\vec{\theta})}{\Psi_i(\vec{\varphi})}|^2,
\end{equation}
where $\Psi_i(\vec{\varphi})$ are the previously found eigenstates with optimized coefficients $\vec{\varphi}$.

This is equivalent to minimizing the objective function
\begin{equation}\label{excited_state_cost_fn_equivalent}
    \hhat{H}_{\mathrm{F},k}^2 = \hhat{H}_\mathrm{F}^2 + \sum_{i=0}^{k-1} \beta_i \kket{\Psi_i(\vec{\varphi})}\bbra{\Psi_i(\vec{\varphi})}.
\end{equation}
The modified cost function in Eq.~(\ref{excited_state_cost_fn_equivalent}) leads to a modified pool operator gradient expression compared to the original expression~\cite{Grimsley_NatCom_2019}. The expression for the gradient of $\langle\hhat{H}_{\mathrm{F},k}^2\rangle$ with respect to pool operator $\hhat{A}_l$ is
\begin{equation}
     \left. \partial \braket{\hhat{H}_{\mathrm{F},k}^2} / \partial \theta_{l}\right|_{\theta_{l}=0} = i\bbra{\Psi(\theta)} [\hhat{A}_l, \hhat{H}_{\mathrm{F},k}^2]\kket{\Psi(\theta)} + 2 \sum_{i=0}^{k-1} \beta_i \bigl( \text{Re} (x_i) \text{Im} (y_i) - \text{Im}(x_i) \text{Re}(y_i) \bigr),
\end{equation}
where $x_i = \iinner{\Psi_i(\vec{\varphi})}{\Psi(\theta)}$, $y_i = \iinner{\Psi_i(\vec{\varphi})}{\hhat{A}_l |\Psi(\theta)}$.

\begin{figure}[H]
    \centering
    \includegraphics[width=0.6\linewidth]{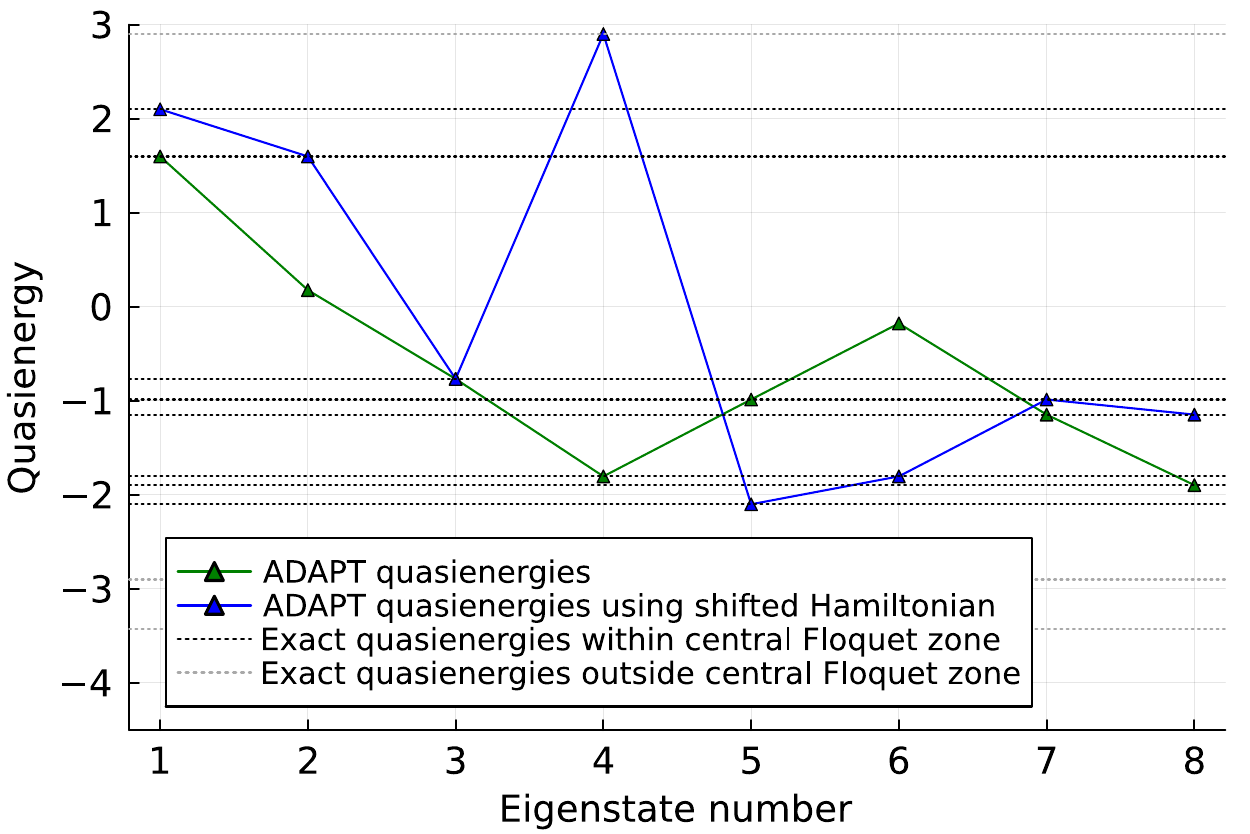}
    \caption{Calculation of quasienergies using the Floquet-ADAPT-VQE algorithm with the modified cost function in Eq.~(\ref{excited_state_cost_fn_equivalent}). All parameters are expressed in units of the characteristic coupling constant $J_0$. We consider a system with $L=3, n_a = 4,  \Omega = 5.0 J_0,  (\bar J_x,\bar J_y,\bar J_z)=(3.7, 2.8, 3.9)J_0, (\delta J_x,\delta J_y,\delta J_z)=(1.9, 1.1, 1.2) J_0$, $\bar B_z=2.9 J_0,\delta B_z=2.7 J_0$. The reference state used here is an equal superposition of all computational basis states, $|+\rangle^{\otimes L_{tot}}$. The shift parameter used to obtain the blue data is $\lambda = 0.6 J_0$.}
    \label{Fig:excited_quasienergies}
\end{figure}

The results of using the cost function Eq.~(\ref{excited_state_cost_fn_equivalent}) in Floquet-ADAPT-VQE to calculate the quasienergy spectra in the central Floquet zone for a system with $L=3, n_a = 4$ are shown in Fig.~\ref{Fig:excited_quasienergies}. For these simulations, we use an operator pool consisting of all possible weight-2 Pauli strings on the total number of qubits, $L + n_a$. The low support of each pool operator lends itself to the strategy of adding multiple disjoint operators at each ADAPT iteration, and these simulations are run using TETRIS-ADAPT-VQE~\cite{Anastasiou_Tetris2024}. The eigenenergies found by Floquet-ADAPT-VQE are depicted in green. Although they are not obtained in an increasing manner, the algorithm is able to find six of the eight exact quasienergies in the central Floquet zone, which are shown as black dotted horizontal lines. The other two quasienergies found by the algorithm correspond to energies of states that are degenerate in the spectrum of the cost function with the squared Floquet Hamiltonian in Eq.~\eqref{excited_state_cost_fn_equivalent}. The degeneracy in the spectrum leads the algorithm to find a linear combination of the two states, where the linear combination may not be an eigenstate of the Floquet Hamiltonian. The problem associated with the degeneracy can be mitigated by adding a small shift to the Floquet Hamiltonian as discussed in Sec.~\ref{sec:Prep_Floq_state}, $\hhat{H}_\mathrm{F} \rightarrow (\hhat{H}_\mathrm{F}-\lambda \hhat{I})$ before squaring it in Eq.~\eqref{excited_state_cost_fn_equivalent}. This leads to the spectrum shown in blue. Note that this is a one-time shift to the Hamiltonian that is not varied. Here, the algorithm finds seven of the eight exact quasienergies in the central Floquet zone, with one corresponding to a quasienergy outside the central Floquet zone.

\section{Floquet-ADAPT-VQE simulations for larger systems}
\begin{figure}[t]
    \centering
    \includegraphics[width=0.7\linewidth]{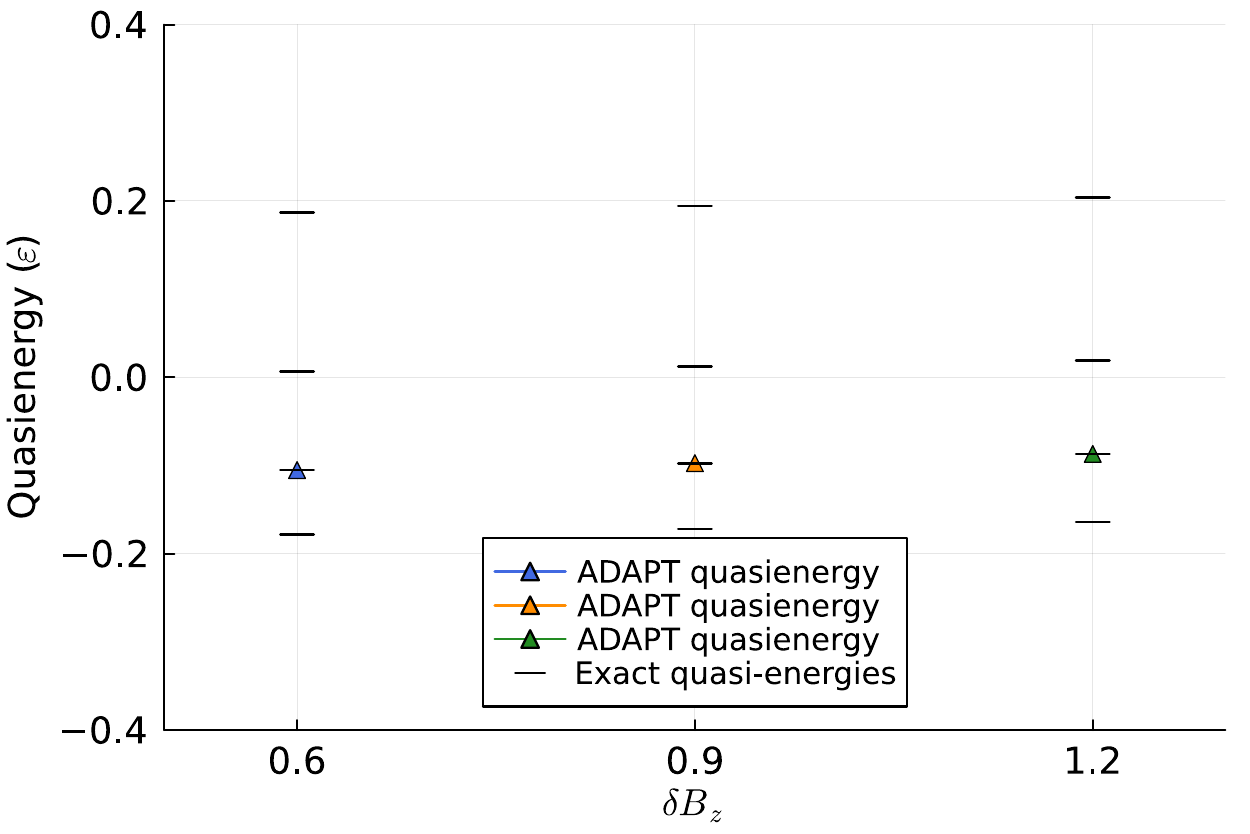}
    \caption{Quasienergy spectrum of a periodically driven XYZ spin chain as a function of the parameter $\delta B_z$ in the cost function. All parameters are expressed in units of the characteristic coupling constant $J_0$. The horizontal markers show the exact quasienergies. The initial state for the Floquet-ADAPT-VQE algorithm is defined as $\kket{\psi_{\text{init}}} = \ket{0}\otimes\ket{\downarrow}^{\otimes L}$. The rest of the parameters are $L=7, n_a=3$, $(\bar{J}_x,\bar{J}_y,\bar{J}_z)=(3.7, 2.8, 3.9) J_0$, $(\delta J_x,\delta J_y,\delta J_z)=(0.0,0.0,0.0) J_0$, $\bar{B}_z= 2.9 J_0$, and $\Omega = 15.0 J_0$. We use an operator pool consisting of all weight-2 Pauli strings on each of the physical and auxiliary registers separately.}
    \label{Fig:Quasienergy_Spectrum_L_7_na_3}
\end{figure}

In this appendix, we present classical simulations of Floquet-ADAPT-VQE for the XYZ model with a driven magnetic field, using a mixed Floquet operator pool composed of all weight-2 Pauli strings acting independently on the auxiliary and physical subspaces. The results of running Floquet-ADAPT-VQE on a system of size $L=7, n_a=3$ qubits, with $\Omega=15.0$ are shown in Fig.~\ref{Fig:Quasienergy_Spectrum_L_7_na_3}. Triangular markers show the quasienergies obtained from ADAPT, and horizontal line markers show the exact quasienergies close to 0.0. We find that Floquet-ADAPT-VQE is able to find one of the quasienergies in the central Floquet zone for each value of $\delta B_z$.  

\section{Some details of detection of time crystal}\label{App:TC_detection}
In this appendix, we elaborate on the possibility of detecting time crystals as briefly discussed in Sec.~\ref{sec:Discussion}. Using Floquet ADAPT-VQE, we prepare an extended Floquet state with quasienergy $\epsilon_1<0$ such that $\kket{\psi(\vec{\theta}_1)}= \hhat{U}(\vec{\theta}_1) \kket{\psi_\mathrm{in}}$ where $\kket{\psi_\mathrm{in}}$ is the initial state. This yields the unitary operator $\hhat U(\vec{\theta}_1)$. We separately prepare a state with a shift parameter $\lambda= \epsilon_1 +\Omega/2$. We repeat this for various sets of disorder couplings. Now we repeat the whole process for as many eigenstates as we can. If we find that the lowest quasienergy is zero in this shifted framework for all considered eigenstates, then we can say that the system would most likely show time crystalline behavior. Furthermore, we can detect the spontaneous breaking of symmetry by taking the expectation values of local operators. Let $\ket{\psi_1}$, $\ket{\psi_2}$ be two Floquet states (their respective extended states are $\kket{\psi_1}$, $\kket{\psi_2}$) with a quasienergy separation of $\Omega/2$. The steady state in the time crystal is a superposition of these two states, given by $\ket{\psi}=(\ket{\psi_1}+ \ket{\psi_2})/\sqrt{2}$. Then stroboscopic expectation value of a local operator $\hat O$ in the steady state is:
\begin{align} 
\bra{\psi(nT)}\hat{O} \ket{\psi(nT)} &= \frac{1}{2}\sum_{j=1}^{2} \bra{\psi_j}\hat{O}\ket{\psi_j} + \mathrm{Re}\big\{ e^{-i n\Omega T/2}\bra{\psi_1}\hat{O}\ket{\psi_2}\big\} \nonumber\\
&=\frac{1}{2}\sum_{j=1}^{2} \bbra{\psi_j}\hhat{O}\kket{\psi_j} + (-1)^n \mathrm{Re} \big\{\bbra{\psi_1}\hhat{O}\kket{\psi_2} \big\}.
\end{align}
Here, $\mathrm{Re}$ denotes the real part of a complex number.
We can calculate the stroboscopic expectation value of the first term of the above equation using a quantum circuit (without the single-qubit $z$-rotations) discussed in the main text. We can calculate the second term as discussed in Refs.~\cite{Ze-Hao2023_Expt_off_diag,Huggins_2020}.

\end{document}